\documentclass[journal]{IEEEtran}

\IEEEoverridecommandlockouts                              

\usepackage{amssymb}
\usepackage{amsmath}
\usepackage{bm}
\usepackage{mathrsfs}
\usepackage{comment} 
\usepackage{algorithmic,algorithm}
\usepackage{shuffle}
\usepackage{graphicx} 
\usepackage{caption}
\usepackage{lipsum}
\usepackage{color}
\usepackage{enumerate}
\usepackage{amsfonts}
\usepackage{subfigure} 
\usepackage{array}
\usepackage{setspace}
\usepackage{multirow}

\DeclareSymbolFont{symbolsC}{U}{pxsyc}{m}{n}
\DeclareMathSymbol{\coloneqq}{\mathrel}{symbolsC}{"42}

\begin{document}

\title{\LARGE \bf
Supervisor Obfuscation Against Covert Actuator Attackers} 
\author{Ruochen Tai, Liyong Lin and Rong Su

%
%
%
\thanks{The research of the project was supported by the Agency for Science, Technology and Research (A*STAR) under its IAF-ICP Programme ICP1900093 and the Schaeffler Hub for Advanced Research at NTU.

The authors are affliated with Nanyang Technological University, Singapore. (Email: ruochen001@e.ntu.edu.sg; llin5@e.ntu.edu.sg; rsu@ntu.edu.sg). (\emph{Corresponding author: Liyong Lin})}
}
\maketitle

\begin{abstract}
This work investigates the problem of synthesizing obfuscated supervisors against covert actuator attackers. For a non-resilient supervisor $S$, for which there exist some covert actuator attackers that are capable of inflicting damage, we propose an algorithm to compute all the obfuscated supervisors, with the requirements that: 1) any obfuscated supervisor $S'$ is resilient against any covert actuator attacker, and 2) the original closed-behavior of the closed-loop system under $S$ is preserved, that is, any obfuscated supervisor $S'$ is control equivalent to the original non-resilient supervisor $S$. We prove that the designed algorithm to synthesize  obfuscated supervisors against covert actuator attack is sound and complete.
\end{abstract}




\section{Introduction}
\label{sec:intro}

With the rapid development of information communication technology, the network has become an indispensable ingredient in the social development, where plenty of transmitted data contain sensitive or even confidential information, inevitably attracting  malicious attacks such as data tampering. 
To defend against those growing, rampant and sophisticated cyber threats, cyber defense strategies, which focus on preventing, detecting and responding to attacks or threats in a timely manner so that infrastructure or information is not tampered with, are essential for most entities to protect sensitive information as well as protect assets, and have attracted more and more attention from both the computer science community and control systems community. 

In this work, we shall investigate the cyber defense strategies at the supervisory control layer, where the system is modeled as discrete-event systems (DES) with discrete state space and event-triggered dynamics. Existing works have proposed plenty of defending strategies, including: 1) synthesis of resilient supervisors \cite{Su2018}-\cite{WLYWL21} such that there does not exist any (covert) sensor-actuator attacker that could induce the plant to the damage state via altering sensor readings and disrupting control signals on those vulnerable observation-command sequence encoded in the supervisor, 2) supervisor obfuscation \cite{Zhu2018}, which computes resilient supervisors that are control equivalent to the original insecure supervisor, that is, on one hand, the original closed-behavior of the closed-loop system is preserved, and on the other hand, the insecure supervisor is obfuscated in the sense that the plant under the new supervisor is not attackable, 3) deploying secure communication channel \cite{WP19}, which guarantees the confidentiality of data to prevent attackers from eavesdropping and intercepting secret information, 4) embedding the mitigation module to disable events that can be defended \cite{YYL20}, 5) designing transition protecting policies against an external intruder, which could delay the event firings, to guarantee that the makespan of the requirement does not drop below a given deadline \cite{HMT21}, 6) synthesizing secret protection strategies such that any event sequence from the initial state that reaches a secret state contains a number of protected events no less than a given threshold \cite{MC21,MC21Conf}, and 7) synthesizing liveness-enforcing supervisors against attacks \cite{YWS22}.
This paper would continue our previous work \cite{Zhu2018} on the study of supervisor obfuscation problem. \cite{Zhu2018} proposes a method to firstly synthesize control equivalent supervisors by reducing it to the Boolean Satisfiability Problem (SAT), and then carry out a verification procedure on each supervisor to check the attackability. However, there are some shortcomings for this approach: 1) it can only compute bounded supervisors by using  SAT solvers; thus, the procedure proposed in \cite{Zhu2018} is  generally incomplete, and 2) the process of enumerating each behavior-preserving supervisor and verifying its attackability causes massive degradation on the performance of the overall algorithm. 
Here we remark that the approach of synthesizing bounded resilient supervisors proposed in \cite{LZS19b,LS20BJ} can also be adopted to address the problem of supervisor obfuscation, by which there is no need to verify the attackability of each control equivalent supervisor. However, it remains a bounded synthesis approach, and thus not complete.

In this work, we propose a sound and complete method to synthesize obfuscated supervisors against covert actuator attackers. The contributions are summarized as follows:
\begin{enumerate}[1.]
\setlength{\itemsep}{3pt}
\setlength{\parsep}{0pt}
\setlength{\parskip}{0pt}
    \item Different from the constraint-based synthesis approach adopted in \cite{Zhu2018}, which can only generate bounded control equivalent supervisors, we provide a new  algorithm to construct a structure, named as behavior-preserving structure, such that it exactly encodes all the control equivalent supervisors.
    \item Instead of verifying the attackability of the control equivalent supervisors one by one, which is adopted in \cite{Zhu2018}, we propose a new approach to directly extract from the behavior-preserving structure all the control equivalent supervisors that are resilient against any covert actuator attacker. We prove that the proposed algorithm for synthesizing obfuscated supervisors against covert actuator attackers is sound and complete. In comparison, the algorithm proposed in \cite{Zhu2018} is sound, but generally incomplete due to the constraint of bounded state size when generating control equivalent supervisors. The same remark in terms of incompleteness also holds for the bounded synthesis approach in \cite{LZS19b,LS20BJ}. 
\end{enumerate}
This paper is organized as follows. In Section \ref{sec:Preliminaries}, we recall the preliminaries which are needed for understanding this paper. In Section \ref{sec:Construction of Behavior-Preserving structure}, we introduce the system architecture and propose a method for constructing a behavior-preserving structure to encode all the control equivalent supervisors. The approach of extracting  resilient supervisors from the behavior-preserving structure is presented in Section \ref{sec:Synthesis of Obfuscated Supervisors Against Covert Actuator Attackers}. Conclusions are drawn in Section \ref{sec:conclusions}. A running example is given throughout the paper. 


\section{Preliminaries}
\label{sec:Preliminaries}
In this section, we introduce some basic notations and terminologies that will be used in this work, mostly following~\cite{WMW10, CL99, HU79}.
Given a finite alphabet $\Sigma$, let $\Sigma^{*}$ be the free monoid over $\Sigma$ with the empty string $\varepsilon$ being the unit element. 
For a string $s$, $|s|$ is defined to be the length of $s$. Given two strings $s, t \in \Sigma^{*}$, we say $s$ is a prefix substring of $t$, written as $s \leq t$, if there exists $u \in \Sigma^{*}$ such that $su = t$, where $su$ denotes the concatenation of $s$ and $u$. 
A language $L \subseteq \Sigma^{*}$ is a set of strings. 
The prefix closure of $L$ is defined as $\overline{L} = \{u \in \Sigma^{*} \mid (\exists v \in L) \, u\leq v\}$. 
If $L = \overline{L}$, then $L$ is prefix-closed. The concatenation of two languages $L_{a}, L_{b} \subseteq \Sigma^{*}$ is defined as $L_{a}L_{b} = \{s_{a}s_{b} \in \Sigma^{*}|s_{a} \in L_{a} \wedge s_{b} \in L_{b}\}$. 
$\mathcal{P}_{j}(s)$ represents the prefix of length $j$, specifically, $\mathcal{P}_{0}(\cdot) = \varepsilon$. 
$s[i]$ denotes the $i$-th element in $s$. $s^{\downarrow}$ denotes the last event in $s$.
The event set $\Sigma$ is partitioned into $\Sigma = \Sigma_{c} \dot{\cup} \Sigma_{uc} = \Sigma_{o} \dot{\cup} \Sigma_{uo}$, where $\Sigma_{c}$ (respectively, $\Sigma_{o}$) and $\Sigma_{uc}$ (respectively, $\Sigma_{uo}$) are defined as the sets of controllable (respectively, observable) and uncontrollable (respectively, unobservable) events, respectively.  As usual, $P_{o}: \Sigma^{*} \rightarrow \Sigma_{o}^{*}$ is the natural projection defined as follows: 1) $P_{o}(\varepsilon) = \varepsilon$, 2) $(\forall \sigma \in \Sigma) \, P_{o}(\sigma) = \sigma$ if $\sigma \in \Sigma_{o}$, otherwise, $P_{o}(\sigma) = \varepsilon$, 3) $(\forall s \in \Sigma^{*}, \sigma \in \Sigma) \, P_{o}(s\sigma) = P_{o}(s)P_{o}(\sigma)$.
We sometimes also write $P_o$ as $P_{\Sigma_o}$, to explicitly illustrate the co-domain $\Sigma_o^*$.

A finite state automaton $G$ over $\Sigma$ is given by a 5-tuple $(Q, \Sigma, \xi, q_{0}, Q_{m})$, where $Q$ is the state set, $\xi: Q \times \Sigma \rightarrow Q$ is the (partial) transition function, $q_{0} \in Q$ is the initial state, and $Q_{m}$ is the set of marker states. 
We write $\xi(q, \sigma)!$ to mean that $\xi(q, \sigma)$ is defined. We define $En_{G}(q) = \{\sigma \in \Sigma|\xi(q, \sigma)!\}$.
$\xi$ is also extended to the (partial) transition function $\xi: Q \times \Sigma^{*} \rightarrow Q$ and the transition function $\xi: 2^{Q} \times \Sigma \rightarrow 2^{Q}$ \cite{WMW10}, where the later is defined as follows: for any $Q' \subseteq Q$ and any $\sigma \in \Sigma$, $\xi(Q', \sigma) = \{q' \in Q|(\exists q \in Q')q' = \xi(q, \sigma)\}$. 
Let $L(G)$ and $L_{m}(G)$ denote the closed-behavior and the marked behavior, respectively. $G$ is said to be marker-reachable if some marker state of $G$ is reachable~\cite{WMW10}, i.e., $L_m(G) \neq \varnothing$. When $Q_{m} = Q$, we shall also write $G = (Q, \Sigma, \xi, q_{0})$ for simplicity. $Ac(G)$ stands for the automaton by taking the ``accessible'' part of $G$ \cite{CL99}, i.e., deleting those states (and the associated transitions) that are not reachable from the initial state.
The ``unobservable reach'' of the state $q \in Q$ under the subset of events $\Sigma' \subseteq \Sigma$ is given by $UR_{G, \Sigma - \Sigma'}(q) := \{q' \in Q|[\exists s \in (\Sigma - \Sigma')^{*}] \, q' = \xi(q,s)\}$.
We shall abuse the notation and define $P_{\Sigma'}(G)$ to be the finite state automaton $(2^{Q} - \{\varnothing\}, \Sigma, \delta, UR_{G, \Sigma - \Sigma'}(q_{0}))$ over $\Sigma$, where $UR_{G, \Sigma - \Sigma'}(q_{0}) \in 2^Q-\{\varnothing\}$ is the initial state, and the (partial) transition function $\delta: (2^{Q} - \{\varnothing\}) \times \Sigma \rightarrow (2^{Q} - \{\varnothing\})$ is defined as follows:
\begin{enumerate}[(1)]
\setlength{\itemsep}{3pt}
\setlength{\parsep}{0pt}
\setlength{\parskip}{0pt}
    \item For any $\varnothing \neq Q' \subseteq Q$ and any $\sigma \in \Sigma'$, if $\xi(Q', \sigma) \neq \varnothing$, then $\delta(Q', \sigma) = UR_{G, \Sigma - \Sigma'}(\xi(Q', \sigma))$, where $UR_{G, \Sigma - \Sigma'}(Q'') = \bigcup\limits_{q \in Q''}UR_{G, \Sigma - \Sigma'}(q)$
    for any $\varnothing \neq Q'' \subseteq Q$.
    \item For any $\varnothing \neq Q' \subseteq Q$ and any $\sigma \in \Sigma - \Sigma'$, if there exists $q \in Q'$ such that $\xi(q, \sigma)!$, then $\delta(Q', \sigma) = Q'$. 
\end{enumerate}

As usual, for any two finite state automata $G_{1} = (Q_{1}, \Sigma_{1}, \xi_{1}, q_{1,0}, Q_{1,m})$ and $G_{2} = (Q_{2}, \Sigma_{2}, \xi_{2}, q_{2,0}, Q_{2,m})$, where $En_{G_{1}}(q) = \{\sigma \in \Sigma_1|\xi_{1}(q, \sigma)!\}$ and $En_{G_{2}}(q) = \{\sigma \in \Sigma_2|\xi_{2}(q, \sigma)!\}$, their synchronous product \cite{CL99} is defined to be  $G_{1}||G_{2} := (Q_{1} \times Q_{2}, \Sigma_{1} \cup \Sigma_{2}, \zeta, (q_{1,0}, q_{2,0}), Q_{1,m} \times Q_{2,m})$, where the (partial) transition function $\zeta$ is defined as follows, for any $(q_{1}, q_{2}) \in Q_{1} \times Q_{2}$ and $\sigma \in \Sigma = \Sigma_1 \cup \Sigma_2$:
\[
\begin{aligned}
& \zeta((q_{1}, q_{2}), \sigma) := \\ & \left\{
\begin{array}{lcl}
(\xi_{1}(q_{1}, \sigma), \xi_{2}(q_{2}, \sigma))  &      & {\rm if} \, {\sigma \in En_{G_{1}}(q_{1}) \cap En_{G_{2}}(q_{2}),}\\
(\xi_{1}(q_{1}, \sigma), q_{2})       &      & {\rm if} \, {\sigma \in En_{G_{1}}(q_{1}) \backslash \Sigma_{2},}\\
(q_{1}, \xi_{2}(q_{2}, \sigma))       &      & {\rm if} \, {\sigma \in En_{G_{2}}(q_{2}) \backslash \Sigma_{1},}\\
{\rm not \, defined}  &      & {\rm otherwise.}
\end{array} \right.
\end{aligned}
\]

\textbf{Notation.} Let $\mathbb{N}$ be the set of nonnegative integers, and $\mathbb{N}^{+}$ the set of positive integers. Let $[m:n] := \{m,m+1,\cdots,n\}$ ($m \in \mathbb{N}, n \in \mathbb{N}$).


\section{Construction of Control-Equivalent Supervisors}
\label{sec:Construction of Behavior-Preserving structure}

In this section, we shall firstly introduce the system architecture under actuator attack. Then, we shall briefly introduce the main idea of our solution methodology. Finally, given the plant and the supervisor, the procedure of constructing the behavior-preserving structure that encodes all the control-equivalent supervisors is presented.

\subsection{Component models}
\label{subsec:Component models}

\begin{figure}[htp]
\begin{center}
\includegraphics[height=2.8cm]{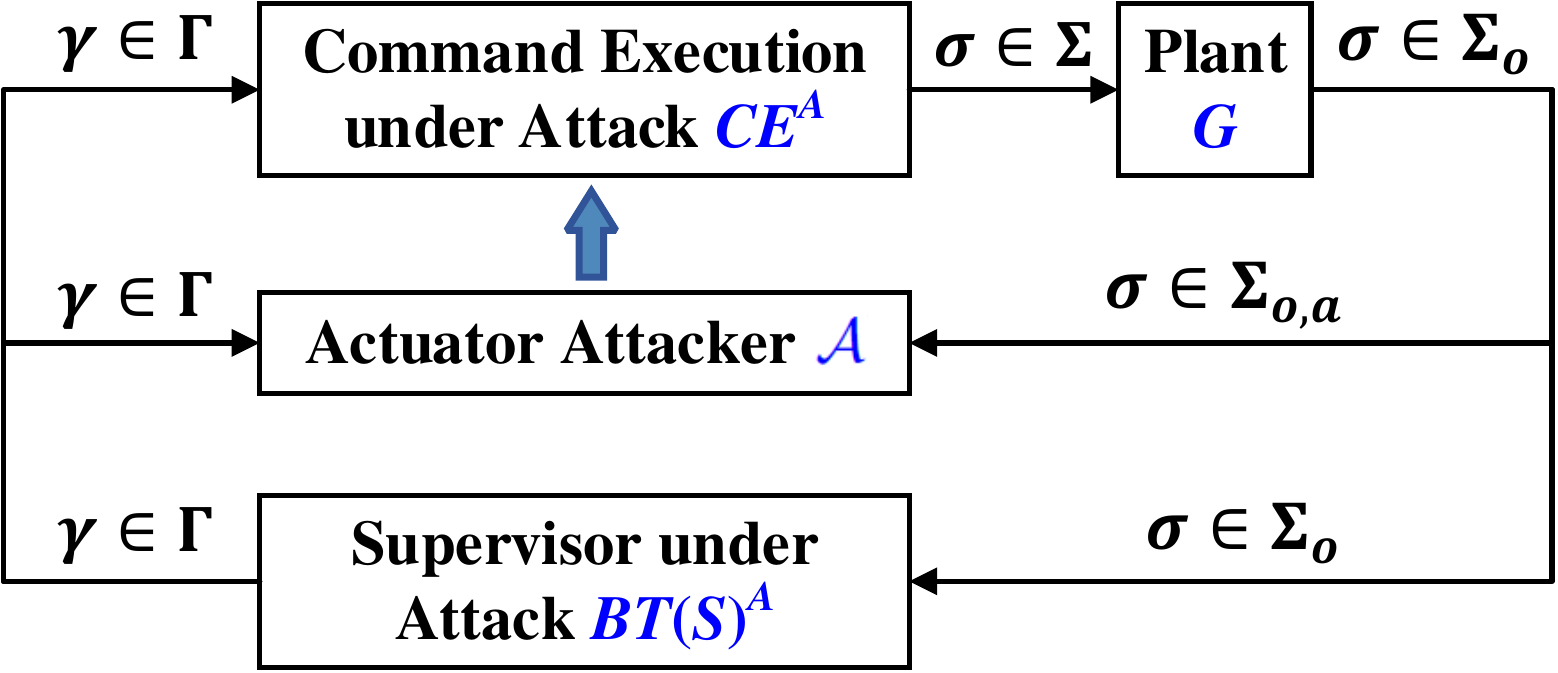}   
\caption{Supervisory control architecture under actuator attack}
\label{fig:architecture}
\end{center}        
\end{figure}

The supervisory control architecture under actuator attack is illustrated in Fig. \ref{fig:architecture}, which consists of the following components: 1) Plant $G$, 2) Supervisor under attack $BT(S)^{A}$, 3) Command execution automaton under attack $CE^{A}$, and 4) Actuator attacker $\mathcal{A}$.
Next, following \cite{LZS19}, we shall briefly explain how we can  model these components as finite state automata.

\subsubsection{Plant}
\label{subsubsec:Plant}

The plant is modeled as a finite state automaton $G = (Q, \Sigma, \xi, q^{init}, Q_{d})$, where $Q_{d}$ is the set of damage states that the actuator attacker targets to induce the plant to reach.

\subsubsection{Supervisor}
\label{subsubsec:Supervisor}

The original supervisor is modeled as a finite state automaton $S = (Q_{s}, \Sigma, \xi_{s}, q_{s}^{init})$ satisfying two constraints: 
\begin{itemize}
\setlength{\itemsep}{3pt}
\setlength{\parsep}{0pt}
\setlength{\parskip}{0pt}
    \item (controllability) For any state $q \in Q_{s}$ and any event $\sigma \in \Sigma_{uc}$, $\xi_{s}(q, \sigma)!$.
    \item (observability) For any state $q \in Q_{s}$ and any event $\sigma \in \Sigma_{uo}$, if $\xi_{s}(q, \sigma)!$, then $\xi_{s}(q, \sigma) = q$.
\end{itemize}
The control command issued by $S$ at state $q \in Q_{s}$ is defined to be $\Gamma(q) = En_{S}(q) = \{\sigma \in \Sigma|\xi_{s}(q,\sigma)!\} \in \Gamma = \{\gamma \subseteq \Sigma|\Sigma_{uc} \subseteq \gamma\}$, where $\Gamma$ is the set of control commands. We assume the supervisor $S$ will immediately issue a control command in $\Gamma$ to the plant whenever an event $\sigma \in \Sigma_{o}$ is received or when the system initiates. Next, based on the supervisor $S$, we construct a bipartite structure to explicitly encode the observation reception and command sending phase. Such a structure is named as bipartite supervisor \cite{LZS19}, which is denoted by $BT(S) = (Q_{bs}, \Sigma_{bs}, \xi_{bs}, q_{bs}^{init})$ and its construction procedure is given as follows:
\begin{enumerate}[1.]
\setlength{\itemsep}{3pt}
\setlength{\parsep}{0pt}
\setlength{\parskip}{0pt}
    \item $Q_{bs} = Q_{s} \cup Q_{s}^{com}$, where $Q_{s}^{com}:= \{q^{com} \mid q \in Q_s\}$, and $q \in Q_{s}$ is a reaction state ready to observe any event in $\Gamma(q)$, and $q^{com} \in Q_{s}^{com}$ is a control state corresponding to $q$, which is ready to issue the control command $\Gamma(q)$.
    \item $\Sigma_{bs} = \Sigma \cup \Gamma$
    \item \begin{enumerate}[a.]
        \setlength{\itemsep}{3pt}
        \setlength{\parsep}{0pt}
        \setlength{\parskip}{0pt}
            \item $(\forall q^{com} \in Q_{s}^{com}) \, \xi_{bs}(q^{com}, \Gamma(q)) = q$.
            \item $(\forall q \in Q_{s})(\forall \sigma \in \Sigma_{uo}) \, \xi_{s}(q, \sigma)! \Rightarrow \xi_{bs}(q, \sigma) = \xi_{s}(q, \sigma) =q$.
            \item $(\forall q \in Q_{s})(\forall \sigma \in \Sigma_{o}) \, \xi_{s}(q, \sigma)! \Rightarrow \xi_{bs}(q, \sigma) = (\xi_{s}(q, \sigma))^{com}$. 
        \end{enumerate}
    \item $q_{bs}^{init} = (q_{s}^{init})^{com}$
\end{enumerate}
The basic idea of constructing a bipartite supervisor is: 1) at any control state $q^{com}$, a control command $\Gamma(q)$ should be issued, which leads to a reaction state $q$ (Case 3.a), 2) at any reaction state $q$, any unobservable event, if defined in $S$, is a self-loop transition (Case 3.b), and any observable event, if defined in $S$, would lead to a control state $(\xi_{s}(q, \sigma))^{com}$ (Case 3.c).
In this work, the set of observable events in $\Sigma$ for the actuator attacker is denoted by $\Sigma_{o,a} \subseteq \Sigma$. All the control commands in $\Gamma$ are observable to the actuator attacker. The set of actuator attackable events is denoted by $\Sigma_{c,a} \subseteq \Sigma_{c}$, i.e., it can enable or disable the execution of events in $\Sigma_{c,a}$ at the plant. 
We assume that $\Sigma_{c,a} \subseteq \Sigma_{o,a}$. 

\textbf{Remark III.1:} The assumptions of $\Sigma_{o,a} \subseteq \Sigma_{o}$ and $\Sigma_{c} \subseteq \Sigma_{o}$, which are imposed in \cite{Zhu2018}, are relaxed in this work. Thus, we consider a more general setup than \cite{Zhu2018}.

Next, we shall encode the attack effects into $BT(S)$ to generate the bipartite supervisor under attack, which is denoted by $BT(S)^{A} = (Q_{bs}^{a}, \Sigma_{bs}^{a}, \xi_{bs}^{a}, q_{bs}^{a,init})$, where: 
\begin{enumerate}[1.]
\setlength{\itemsep}{3pt}
\setlength{\parsep}{0pt}
\setlength{\parskip}{0pt}
    \item $Q_{bs}^{a} = Q_{bs} \cup \{q^{detect}\}$
    \item $\Sigma_{bs}^{a} = \Sigma \cup \Gamma$
    \item \begin{enumerate}[a.]
        \setlength{\itemsep}{3pt}
        \setlength{\parsep}{0pt}
        \setlength{\parskip}{0pt}
        \item $(\forall q, q' \in Q_{bs}^{a})(\forall \sigma \in \Sigma \cup \Gamma) \xi_{bs}(q, \sigma) = q' \Rightarrow \xi_{bs}^{a}(q, \sigma) = q'$
        \item $(\forall q \in Q_{s})(\forall \sigma \in \Sigma_{c,a} \cap \Sigma_{uo}) \neg\xi_{bs}(q, \sigma)! \Rightarrow \xi_{bs}^{a}(q, \sigma) = q$ 
        \item $(\forall q \in Q_{s})(\forall \sigma \in \Sigma_{o}) \neg\xi_{bs}(q, \sigma)! \Rightarrow \xi_{bs}^{a}(q, \sigma) = q^{detect}$ 
    \end{enumerate}
    \item $q_{bs}^{a,init} = q_{bs}^{init}$
\end{enumerate}
Step 3.a retains all the transitions originally defined in $BT(S)$. In Step 3.b, for any reaction state $q \in Q_{s}$, the transitions labelled by unobservable and attackable events in $\Sigma_{c,a} \cap \Sigma_{uo}$, which are not originally defined at the state $q$ in $BT(S)$, are added. In Step 3.c, for any reaction state $q \in Q_{s}$, the transitions labelled by observable events, which are not originally defined at the state $q$ in $BT(S)$, would lead to the newly added state $q^{detect}$, with the interpretation that the supervisor has received some observation that should not have occurred based on the supervisor structure, i.e., the actuator attacker is detected, and then the system operation would be halted.

\subsubsection{Command execution automaton}
\label{subsubsec:command execution automaton}

To explicitly encode the phase from receiving a control command in $\Gamma$ to executing an event in $\Sigma$ at the plant, we construct the command execution automaton $CE = (Q_{ce}, \Sigma_{ce}, \xi_{ce}, q_{ce}^{init})$, where $Q_{ce} = \{q^{\gamma}|\gamma \in \Gamma\} \cup \{q_{ce}^{init}\}$, $\Sigma_{ce} = \Gamma \cup \Sigma$ and the (partial) transition function $\xi_{ce}: Q_{ce} \times \Sigma_{ce} \rightarrow Q_{ce}$ is defined as follows: 
\begin{enumerate}[1.]
\setlength{\itemsep}{3pt}
\setlength{\parsep}{0pt}
\setlength{\parskip}{0pt}
    \item $(\forall \gamma \in \Gamma) \xi_{ce}(q_{ce}^{init}, \gamma) = q^{\gamma}$
    \item $(\forall \gamma \in \Gamma)(\forall \sigma \in \gamma \cap \Sigma_{o}) \xi_{ce}(q^{\gamma}, \sigma) = q_{ce}^{init}$.
    \item $(\forall \gamma \in \Gamma)(\forall \sigma \in \gamma \cap \Sigma_{uo}) \xi_{ce}(q^{\gamma}, \sigma) = q^{\gamma}$.
\end{enumerate}
Next, we encode the attack effects into $CE$ to generate the command execution automaton under attack, which is denoted by $CE^{A} = (Q_{ce}^{a}, \Sigma \cup \Gamma, \xi_{ce}^{a}, q_{ce}^{a,init})$, where $Q_{ce}^{a} = Q_{ce}$, $q_{ce}^{a,init} = q_{ce}^{init}$ and the (partial) transition function $\xi_{ce}^{a}: Q_{ce}^{a} \times (\Sigma \cup \Gamma) \rightarrow Q_{ce}^{a}$ is defined as follows:
\begin{enumerate}[1.]
\setlength{\itemsep}{3pt}
\setlength{\parsep}{0pt}
\setlength{\parskip}{0pt}
    \item $(\forall q, q' \in Q_{ce}^{a})(\forall \sigma \in \Sigma \cup \Gamma) \xi_{ce}(q, \sigma) = q' \Rightarrow \xi_{ce}^{a}(q, \sigma) = q'$ 
    \item $(\forall \gamma \in \Gamma)(\forall \sigma \in \Sigma_{c,a} \cap \Sigma_{o}) \neg\xi_{ce}(q^{\gamma}, \sigma)! \Rightarrow \xi_{ce}^{a}(q^{\gamma}, \sigma) = q_{ce}^{a,init}$
    \item $(\forall \gamma \in \Gamma)(\forall \sigma \in \Sigma_{c,a} \cap \Sigma_{uo}) \neg\xi_{ce}(q^{\gamma}, \sigma)! \Rightarrow \xi_{ce}^{a}(q^{\gamma}, \sigma) = q^{\gamma}$
\end{enumerate}
Case 1 retains all the transitions originally defined in $CE$. In Case 2 and Case 3, the attack effects are encoded: for any state $q^{\gamma}$, the transitions labelled by attackable events, which are not originally defined at the state $q^{\gamma}$ in $CE$, are added, where the observable events would lead to the initial state $q_{ce}^{init}$ (Case 2), and the unobservable events would lead to self-loop transitions (Case 3).

\subsubsection{Actuator attacker}
\label{subsubsec:Actuator attacker}

The actuator attacker is modeled by a finite state automaton $\mathcal{A} = (Q_{a}, \Sigma_{a}, \xi_{a}, q_{a}^{init})$, where $\Sigma_{a} = \Sigma \cup \Gamma$. There are two conditions that need to be satisfied:
\begin{itemize}
\setlength{\itemsep}{3pt}
\setlength{\parsep}{0pt}
\setlength{\parskip}{0pt}
    \item ($\mathcal{A}$-controllability) For any state $q \in Q_{a}$ and any event $\sigma \in \Sigma_{a,uc} := \Sigma_{a} - \Sigma_{c,a}$, $\xi_{a}(q, \sigma)!$ 
    \item ($\mathcal{A}$-observability) For any state $q \in Q_{a}$ and any event $\sigma \in \Sigma_{a,uo} := \Sigma_{a} - (\Sigma_{o,a} \cup \Gamma)$, if $\xi_{a}(q, \sigma)$!, then $\xi_{a}(q, \sigma) = q$.
\end{itemize}
$\mathcal{A}$-controllability states that the actuator attacker can only disable events in $\Sigma_{c,a}$. $\mathcal{A}$-observability states that the actuator attacker can only make a state change after observing an event in $\Sigma_{o,a} \cup \Gamma$. In the following text, we shall refer to $(\Sigma_{o,a}, \Sigma_{c,a})$ as the attack constraint, and $\mathscr{C}_{ac} = (\Sigma_{c,a}, \Sigma_{o,a} \cup \Gamma)$ as the attacker's control constraint.

Based on the above-constructed component models, including the plant $G$, the bipartite supervisor under attack $BT(S)^{A}$, the command execution automaton under attack $CE^{A}$ and the actuator attacker $\mathcal{A}$, the closed-loop system under attack is denoted by $CLS^{A} = G||CE^{A}||BT(S)^{A}||\mathcal{A} = (Q_{b}^{a}, \Sigma_{b}^{a}, \xi_{b}^{a}, q_{b}^{a,init}, Q_{b,m}^{a})$.

\textbf{Definition III.1 (Covertness):} Given $G$, $BT(S)^{A}$ and $CE^{A}$, an actuator attacker $\mathcal{A}$ is said to be covert against the supervisor $S$ w.r.t. the  attack constraint $(\Sigma_{o,a}, \Sigma_{c,a})$ if any state in $\{(q_{g},q_{ce}^{a},q_{bs}^{a}, q_{a}) \in Q_{b}^{a}| q_{bs}^{a} = q^{detect}\}$ is not reachable in $CLS^{A}$.

\textbf{Definition III.2 (Damage-reachable):} Given $G$, $BT(S)^{A}$ and $CE^{A}$, an actuator attacker $\mathcal{A}$ is said to be damage-reachable against the supervisor $S$ w.r.t. the  attack constraint $(\Sigma_{o,a}, \Sigma_{c,a})$ if $L_{m}(CLS^{A}) \neq \varnothing$.

\textbf{Definition III.3 (Resilience):} Given $G$, a supervisor $S$ is said to be resilient if there does not exist any covert and damage-reachable actuator attacker $\mathcal{A}$ against $S$ w.r.t. the attack constraint $(\Sigma_{o,a}, \Sigma_{c,a})$. 

In this work, we assume that the original supervisor $S$ is not resilient for the plant $G$, i.e., there exists a covert and damage-reachable actuator attacker $\mathcal{A}$ against $S$ w.r.t. the attack constraint $(\Sigma_{o,a}, \Sigma_{c,a})$. 

\textbf{Definition III.4 (Control equivalence):} Given $G$ and $S$, a supervisor $S'$ (bipartite supervisor $BT(S')$, respectively) is said to be control equivalent to $S$ ($BT(S)$, respectively) if $L(G||S) = L(G||S')$ ($P_{\Sigma}(L(G||CE||BT(S))) = P_{\Sigma}(L(G||CE||BT(S')))$, respectively)\footnote{By construction, $CE$ indeed encodes all the bipartite supervisors, thus, $L(BT(S)) \subseteq L(CE)$ and $L(BT(S')) \subseteq L(CE)$, implying that $L(G||CE||BT(S)) = L(G||BT(S))$ and $L(G||CE||BT(S')) = L(G||BT(S'))$. Hence, the control equivalence could also be defined as $P_{\Sigma}(L(G||BT(S))) = P_{\Sigma}(L(G||BT(S')))$.}.

With the above definitions, we are ready to introduce the problem to be solved in this work.

\textbf{Problem 1:} Given $G$ and $S$, find a structure that encodes the set of all the resilient supervisors that are control equivalent to $S$.

\textbf{Problem 2:} Given $G$ and $S$, compute a resilient supervisor that is control equivalent to $S$.

\textbf{Remark III.2:} By extracting one supervisor out of the structure that encodes the set of all the resilient supervisors that are control equivalent to $S$, if we could solve \textbf{Problem 1}, then we also solve \textbf{Problem 2}. Thus, in this work, we mainly focus on \textbf{Problem 1}. Later in Section \ref{subsec:Generation of Obfuscated Supervisors Against Covert Actuator Attackers}, we will show how the extraction can be easily carried out.

\begin{figure}[htbp]
\centering
\subfigure[]{
\begin{minipage}[t]{0.4\linewidth}
\centering
\includegraphics[height=0.6in]{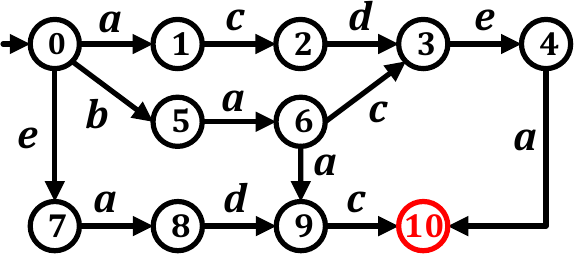}
\end{minipage}
}
\subfigure[]{
\begin{minipage}[t]{0.4\linewidth}
\centering
\includegraphics[height=0.3in]{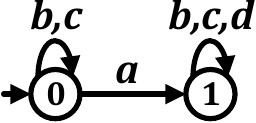}
\end{minipage}
}

\centering
\caption{(a) Plant $G$. (b) Supervisor $S$.}
\label{fig:G_S}
\end{figure}

\begin{figure}[htbp]
\centering
\subfigure[]{
\begin{minipage}[t]{0.4\linewidth}
\centering
\includegraphics[height=0.8in]{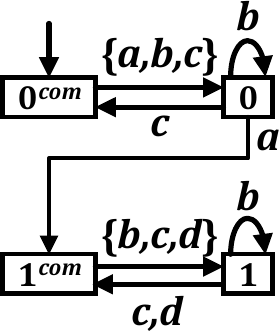}
\end{minipage}
}
\subfigure[]{
\begin{minipage}[t]{0.52\linewidth}
\centering
\includegraphics[height=0.8in]{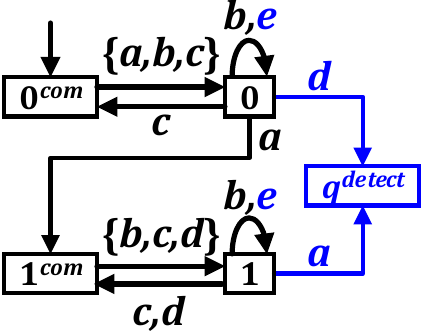}
\end{minipage}
}

\centering
\caption{(a) $BT(S)$. (b) $BT(S)^{A}$.}
\label{fig:BTS_BTSA}
\end{figure}

\begin{figure}[htbp]
\centering
\subfigure[]{
\begin{minipage}[t]{0.4\linewidth}
\centering
\includegraphics[height=1in]{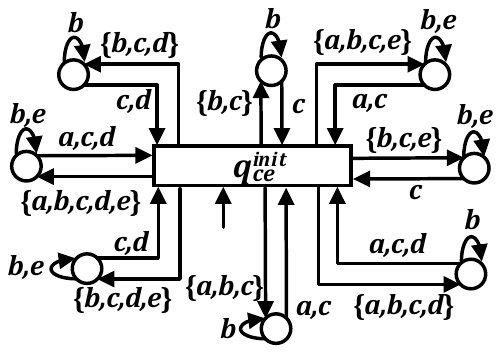}
\end{minipage}
}
\subfigure[]{
\begin{minipage}[t]{0.4\linewidth}
\centering
\includegraphics[height=1in]{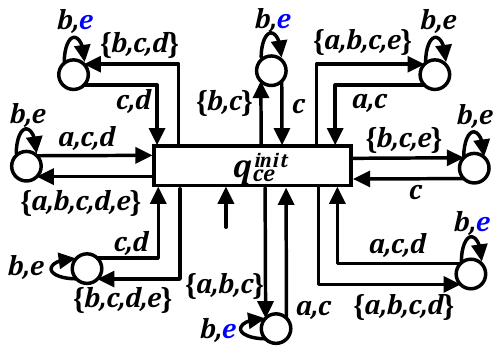}
\end{minipage}
}

\centering
\caption{(a) $CE$. (b) $CE^{A}$.}
\label{fig:CE_CEA}
\end{figure}

\textbf{Example III.1} Consider the plant $G$ and supervisor $S$ shown in Fig. \ref{fig:G_S}. $\Sigma = \{a,b,c,d,e\}$. $\Sigma_{o} = \{a,c,d\}$. $\Sigma_{uo} = \{b,e\}$. $\Sigma_{c} = \{a,d,e\}$. $\Sigma_{uc} = \{b,c\}$. $\Sigma_{o,a} = \{b,c,d,e\}$. $\Sigma_{c,a} = \{e\}$. The damage state is state 10, i.e., $Q_{d} = \{10\}$. We have $L(G||S) = \overline{\{acd,bac\}}$.
Based on the above model constructions, bipartite supervisor $BT(S)$, bipartite supervisor under attack $BT(S)^{A}$, command execution automaton $CE$ and command execution automaton under attack $CE^{A}$ are illustrated in Fig. \ref{fig:BTS_BTSA} and Fig. \ref{fig:CE_CEA}, where the difference between $BT(S)$ and $BT(S)^{A}$, and $CE$ and $CE^{A}$ are marked in blue. It can be checked that $S$ is not resilient for $G$ as there exist covert and damage-reachable attackers. For example, an attacker could implement enablement attack to enable the execution of event $e$ after observing that $S$ issues the initial control command $\{a,b,c\}$. Then $G$ transits to state 7. Since $e$ is unobservable, the command $\{a,b,c\}$ would be reused and event $a$ is executed and observed by $S$, which triggers the sending of command $\{b,c,d\}$. After that, event $d$ is executed and triggers the sending of command $\{b,c,d\}$, resulting in the execution of event $c$ and the damage state is reached. 

\subsection{Main idea}
\label{subsec:Main idea}
\begin{figure}[htbp]
\begin{center}
\includegraphics[height=6.7cm]{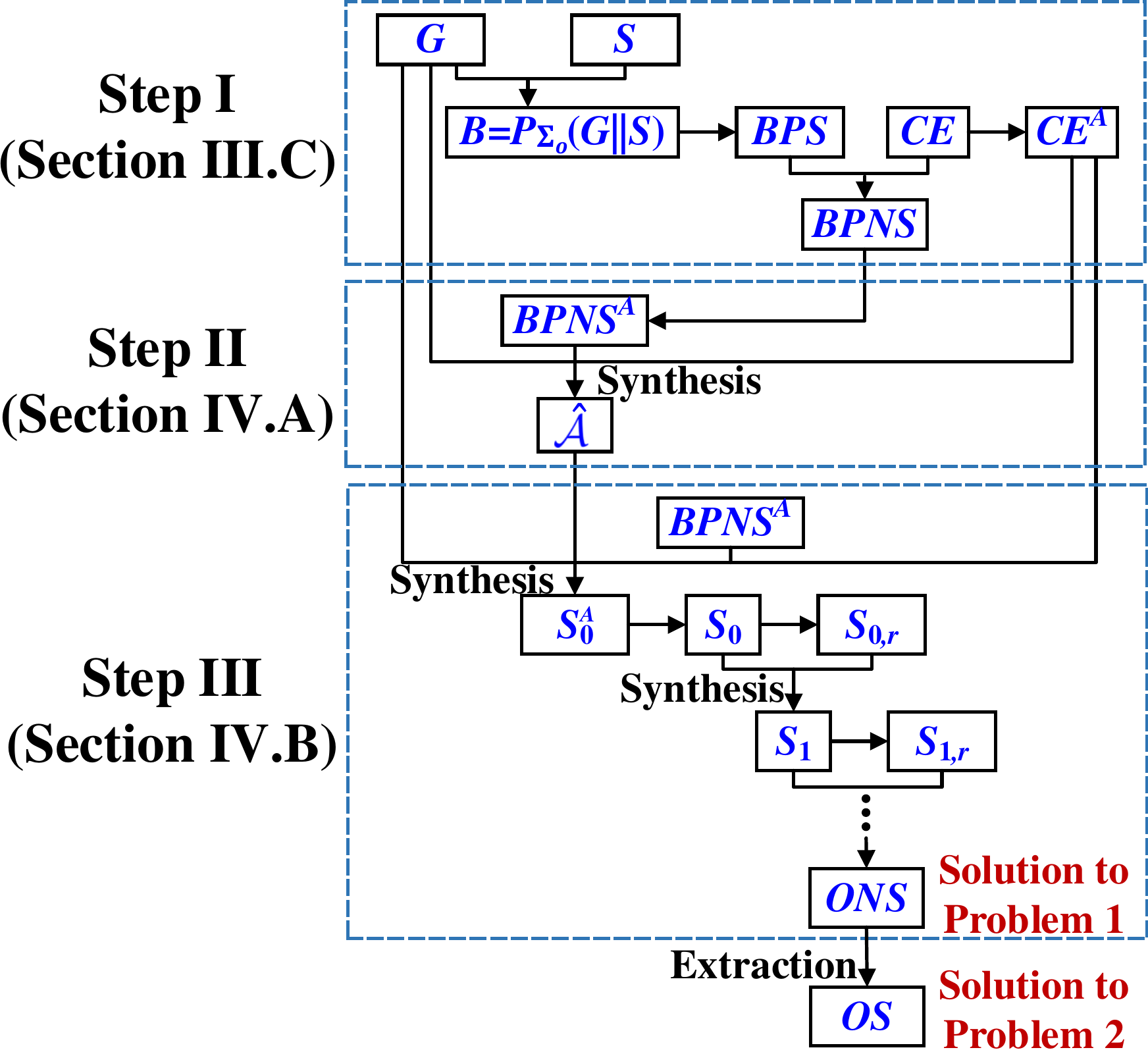}   
\caption{The procedure of the proposed solution methodology}
\label{fig:Solution_Methodology}
\end{center}        
\end{figure}

Before we delve into the details of our proposed approach for supervisor obfuscation, we shall present the high-level idea of the solution methodology, as illustrated in Fig. \ref{fig:Solution_Methodology}. Firstly, at Step I (Section \ref{subsec:Generation of control equivalent bipartite supervisors}), based on $G$, $S$ and $CE$, we shall construct the behavior-preserving structure $BPNS$ to exactly encode all the control equivalent bipartite supervisors. Then, at At Step II (Section \ref{subsec:Synthesis of covert actuator attackers for control equivalent bipartite supervisors}), based on $G$, $CE^{A}$ and $BPNS^{A}$, which is the version of $BPNS$ under attack, we shall synthesize $\hat{\mathcal{A}}$ which encodes all the damage strings that could be taken use of by covert and damage-reachable actuator attackers for control equivalent bipartite supervisors encoded in $BPNS$. Finally, at Step III (Section \ref{subsec:Generation of Obfuscated Supervisors Against Covert Actuator Attackers}), based on $G$, $CE^{A}$, $BPNS^{A}$ and $\hat{\mathcal{A}}$, we shall carry out an iterative synthesis to generate $ONS$, which is the solution to \textbf{Problem 1} that exactly encodes all the resilient and control equivalent supervisors. Then we could extract from $ONS$ a resilient and control equivalent supervisor $OS$, which is the solution to \textbf{Problem 2}. 

\subsection{Generation of control equivalent bipartite supervisors}
\label{subsec:Generation of control equivalent bipartite supervisors}

In this part, we shall introduce the procedure to construct the behavior-preserving structure, where all the bipartite supervisors that are control equivalent to the bipartite supervisor $BT(S)$ are included. The construction procedure consists of the following three steps:

\textbf{Step 1}: Firstly, to encode all the control equivalent supervisors, we need to know the closed-behavior of the closed-loop system under the supervisor $S$ in the absence of attack, thus, we compute $G||S$. Then, since for any supervisor $S'$, we have $L(BT(S')) \subseteq \overline{(\Gamma\Sigma_{uo}^{*}\Sigma_{o})^{*}}$ and any unobservable event defined in $BT(S')$ is a self-loop transition, we shall compute a subset construction $B = P_{\Sigma_{o}}(G||S) = (Q_{b}, \Sigma, \xi_{b}, q_{b}^{init})$, where $|Q_{b}| \leq 2^{|Q| \times |Q_{s}|}$. In fact, $B$ can be regarded as a structure built upon the observer of $G||S$ by adding the self-loop transitions labelled by the unobservable events that could possibly occur at each state of the observer of $G||S$. In addition, it can be checked that, for any $(q_{g}, q_{s}), (q_{g}', q_{s}') \in q \in Q_{b}$, we have $q_{s} = q_{s}'$ since all the unobservable events in $S$ are self-loop transitions. Thus, for any $q, q' \in Q_{b}$ and any $\sigma \in \Sigma_{o}$ such that $\xi_{b}(q, \sigma) = q'$, where $q_{s}$ ($q_{s}'$, respectively) is the supervisor state in the state $q$ ($q'$, respectively), we know that 1) $En_{B}(q')$ contains all the events that could happen at the plant $G$ when the supervisor $S$ issues the corresponding control command at the state $q_{s}'$, and 2) the state of the supervisor $S$ would transit from $q_{s}$ to $q_{s}'$ upon the observation of $\sigma$. Henceforth, we can use $En_{B}(q')$ as the criterion to determine all the possible control commands w.r.t. each observation such that the generated bipartite supervisor $BT(S')$ are control equivalent to the original one, i.e., $BT(S)$. In the next step, we shall explain how to realize this procedure.

\begin{figure}[htbp]
\begin{center}
\includegraphics[height=0.75cm]{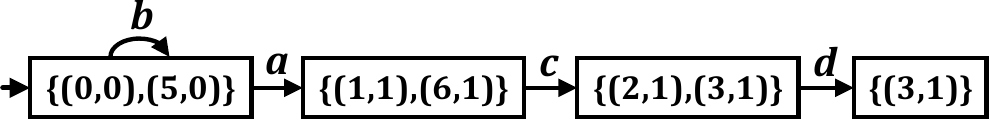}   
\caption{The computed automaton $B$}
\label{fig:B}
\end{center}        
\end{figure}

\textbf{Example III.2} Given $G$ and $S$ shown in Fig. \ref{fig:G_S}, the automaton $B$ is shown in Fig. \ref{fig:B}.

\textbf{Step 2}: Based on $B = P_{\Sigma_{o}}(G||S)$, we shall generate a bipartite structure similar to $BT(S)$, where upon each observation in $P_{\Sigma_{o}}(G||S)$, we add all the possible control commands, under which the closed-behavior of the closed-loop system $L(G||S)$ is preserved. Such a structure is named as bipartite behavior-preserving structure, denoted by $BPS = (Q_{bps}, \Sigma_{bps}, \xi_{bps}, q_{bps}^{init})$, where 
\begin{enumerate}[1.]
\setlength{\itemsep}{3pt}
\setlength{\parsep}{0pt}
\setlength{\parskip}{0pt}
    \item $Q_{bps} = Q_{b} \cup Q_{b}^{com} \cup \{q^{dump}\}$, where $Q_{b}^{com} = \{q^{com}|q \in Q_{b}\}$
    \item $\Sigma_{bps} = \Sigma \cup \Gamma$
    \item \begin{enumerate}[a.]
        \setlength{\itemsep}{3pt}
        \setlength{\parsep}{0pt}
        \setlength{\parskip}{0pt}
        \item $(\forall q \in Q_{b})(\forall \gamma \in \Gamma)\mathcal{C}_{1} \wedge \mathcal{C}_{2} \Rightarrow \xi_{bps}(q^{com}, \gamma) = q$, where
        \begin{enumerate}[i.]
        \setlength{\itemsep}{3pt}
        \setlength{\parsep}{0pt}
        \setlength{\parskip}{0pt}
            \item $\mathcal{C}_{1} := En_{B}(q) \subseteq \gamma$
            \item $\mathcal{C}_{2} := (\forall (q_{g},q_{s}) \in q)En_{G}(q_{g}) \cap \gamma \subseteq En_{B}(q)$
        \end{enumerate} 
        \item $(\forall q \in Q_{b})(\forall \sigma \in \Sigma_{uo})\xi_{b}(q, \sigma)! \Rightarrow \xi_{bps}(q, \sigma) = q$
        \item $(\forall q \in Q_{b})(\forall \sigma \in \Sigma_{o})\xi_{b}(q, \sigma)! \Rightarrow \xi_{bps}(q, \sigma) = (\xi_{b}(q, \sigma))^{com}$
        \item $(\forall q \in Q_{b})(\forall \sigma \in \Sigma_{uo})\neg\xi_{b}(q, \sigma)! \Rightarrow \xi_{bps}(q, \sigma) = q$
        \item $(\forall q \in Q_{b})(\forall \sigma \in \Sigma_{o})\neg\xi_{b}(q, \sigma)! \Rightarrow \xi_{bps}(q, \sigma) = q^{dump}$
        \item $(\forall \sigma \in \Sigma \cup \Gamma)\xi_{bps}(q^{dump}, \sigma) = q^{dump}$
    \end{enumerate}
    \item $q_{bps}^{init} = (q_{b}^{init})^{com}$
\end{enumerate}
In the state set $Q_{bps}$, any state $q^{com} \in Q_{b}^{com}$ is a control state, which is ready to issue the control command, and any state $q$ in $Q_{b}$ is a reaction state, which is ready to receive an observation. After a control command is issued at a control state $q^{com}$, $BPS$ would transit to a reaction state $q$. The state $q^{dump}$ denotes the situation when an event $\sigma \in \Sigma_{o}$, which is not originally defined at the state $q \in Q_{b}$ in $B = P_{\Sigma_{o}}(G||S)$, occurs at the state $q$ in $BPS$. The initial state of $BPS$ is thus the initial control state, denoted by $q_{bps}^{init} = (q_{b}^{init})^{com}$. The definition of the (partial) transition function $\xi_{bps}$ is given in Step 3. Case 3.a adds the control commands that can be issued at any control state $q^{com}$ and the criterion for adding a control command $\gamma \in \Gamma$ at the state $q^{com}$ is: 1) The sending of $\gamma$ should make sure that all the events in $En_{B}(q)$ would occur at the plant $G$ once $\gamma$ is received, denoted by the condition $\mathcal{C}_{1} := En_{B}(q) \subseteq \gamma$; 2) According to the way of constructing $B = P_{\Sigma_{o}}(G||S)$, any state $q \in Q_{b}$ such that $(\exists t \in \Sigma_{o}^{*})\xi_{b}(q_{b}^{init}, t) = q$ already contains all the possible estimated states of the plant $G$ w.r.t. the observation sequence $t$. Henceforth, for any possible plant state $q_{g}$ in the state $q$, the sending of $\gamma$ should make sure that any event that might be executed at the state $q_{g}$ under $\gamma$ would not go beyond $En_{B}(q)$, denoted by the condition $\mathcal{C}_{2} := (\forall (q_{g},q_{s}) \in q)En_{G}(q_{g}) \cap \gamma \subseteq En_{B}(q)$. The conditions $\mathcal{C}_{1}$ and $\mathcal{C}_{2}$ together enforce that at the control state $q^{com}$, any control command $\gamma$ satisfying these two conditions would enable the plant to execute exactly those events in $En_{B}(q)$. Case 3.b and Case 3.c retains all the transitions originally defined in $B$. Next, we shall explain why we add Case 3.d - Case 3.e. Our goal is to construct a structure to include all the bipartite supervisors that are control equivalent to $BT(S)$, where for any supervisor $S' = (Q_{s'}, \Sigma, \xi_{s'}, q_{s'}^{init})$, at any reaction state $q \in Q_{s'}$ of its bipartite version $BT(S')$, all the events in the control command issued at the state $q^{com}$ should be defined. Since Case 3.a has already added all the possible control commands that ensure control equivalence, our basic idea is: 
\begin{enumerate}[1.]
\setlength{\itemsep}{3pt}
\setlength{\parsep}{0pt}
\setlength{\parskip}{0pt}
    \item Firstly, for any reaction state $q \in Q_{b}$ in $BPS$, we shall carry out Case 3.d and Case 3.e to complete all the transitions labelled by events in $\Sigma$ that are not originally defined at the state $q \in Q_{b}$ in $B$. The completed unobservable events would lead to self-loop transitions and completed observable events would result in transitions to the state $q^{dump}$, where any control command is defined, since these completed observable events would not occur at all under the control commands defined at the control state $q^{com}$, and thus the control equivalence would not be violated no matter which command is issued at the state $q^{dump}$.
    \item Then we use $CE$ to refine the above-constructed structure to get all bipartite control equivalent supervisors, which would be done in the later \textbf{Step 3}. 
\end{enumerate}
In Case 3.f, since $BPS$ has transited from some state $q \in Q_{b}$ to the state $q^{dump}$, i.e., some observable event that would never occur under the control of the command issued at the state $q^{com}$ happens, we can safely add the transitions labelled by all the events in $\Sigma \cup \Gamma$ at the state $q^{dump}$ as now the closed-behavior of the closed-loop system would not be affected no matter which control command the supervisor issues.

\textbf{Example III.3} Based on $B$ shown in Fig. \ref{fig:B}, the constructed $BPS$ is illustrated in Fig. \ref{fig:B}. We shall briefly explain the construction procedure by taking two states as instances. At the initial control state $\{(0,0),(5,0)\}^{com}$, 1) according to $\mathcal{C}_{1}$ of Case 3.a, we have $En_{B}(\{(0,0),(5,0)\}) = \{a,b\} \subseteq \gamma$, 2) according to $\mathcal{C}_{2}$ of Case 3.a, we have $En_{G}(0) = \{a,b,e\} \cap \gamma \subseteq En_{B}(\{(0,0),(5,0)\}) = \{a,b\}$, which implies that event $e$ should not be contained in any command, and $En_{G}(5) = \{a\} \cap \gamma \subseteq En_{B}(\{(0,0),(5,0)\}) = \{a,b\}$. Thus, the control commands satisfying $\mathcal{C}_{1}$ and $\mathcal{C}_{2}$ are $\{a,b,c\}$ and $\{a,b,c,d\}$, as uncontrollable events $b$ and $c$ are always contained in any command. Hence, there are two transitions labelled by $\{a,b,c\}$ and $\{a,b,c,d\}$ from the initial state to the reaction state $\{(0,0),(5,0)\}$. This means that at the initial state, to ensure the control equivalence, a supervisor could only issue command $\{a,b,c\}$ or $\{a,b,c,d\}$. At the state $\{(0,0),(5,0)\}$, according to Case 3.b and Case 3.c, there are two transitions labelled by event $b$, which is a self-loop, and event $a$, which leads to the state $(\xi_{b}(\{(0,0),(5,0)\}, a)^{com} = \{(1,1),(6,1)\}^{com}$. According to Case 3.d, a transition labelled by unobservable event $e$ is added at the state $\{(0,0),(5,0)\}$. According to Case 3.e, two transitions labelled by observable events $c$ and $d$ are added at the state $\{(0,0),(5,0)\}$, which lead to the state $q^{dump}$. It can be checked that event $c$ and $d$ would not occur at all under the initial command $\{a,b,c\}$ and $\{a,b,c,d\}$. 

\begin{figure}[htbp]
\begin{center}
\includegraphics[height=3.5cm]{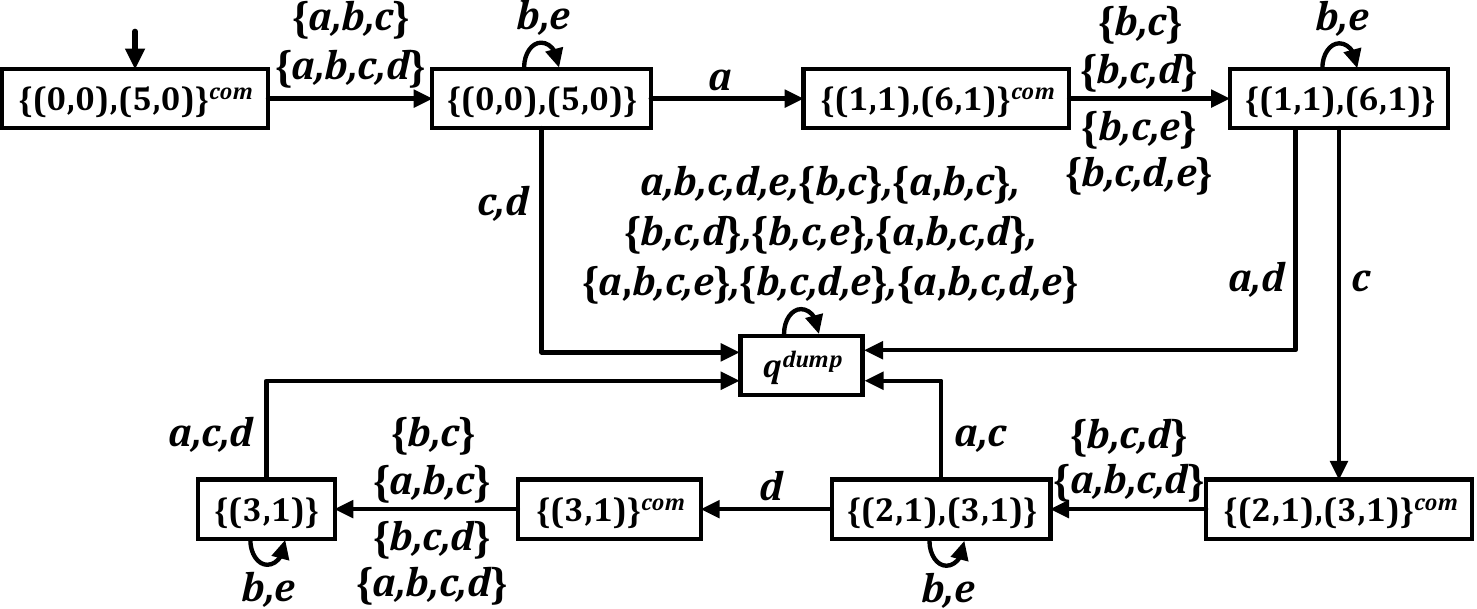}   
\caption{Bipartite behavior-preserving structure $BPS$}
\label{fig:BPS}
\end{center}        
\end{figure}

\textbf{Step 3}: To obtain the final desired structure which contains all the bipartite supervisors that are control equivalent to $BT(S)$, we shall carry out the refinement on $BPS$ by computing the synchronous product of $BPS$ and $CE$. Intuitively speaking, the reason why we could use $CE$ are: 1) $CE$ already encodes all the bipartite supervisors\footnote{This is because: 1) at the state $q_{ce}^{init}$, any control command $\gamma \in \Gamma$ is defined and would lead to the state $q^{\gamma}$, and 2) at any state $q^{\gamma} \in Q_{ce}$, only events in $\gamma$ are defined, and any unobservable event in $\gamma$ would lead to a self-loop transition and any observable event in $\gamma$ would lead to a transition back to the initial state $q_{ce}^{init}$.}, and 2) the structure of $CE$ could ensure that the automaton computed by synchronous product still maintains the structure similar to that of a bipartite supervisor.
We shall name this structure as bipartite behavior-preserving command-nondeterministic\footnote{We note that this structure is a deterministic automaton, but command non-deterministic in the sense that at each control state, more than two different control commands may be issued.} supervisor, denoted by
$BPNS = BPS || CE = (Q_{bpns}, \Sigma \cup \Gamma, \xi_{bpns}, q_{bpns}^{init})$, where $Q_{bpns} = (Q_{b} \cup Q_{b}^{com} \cup \{q^{dump}\}) \times Q_{ce} = (Q_{b} \cup Q_{b}^{com} \cup \{q^{dump}\}) \times (\{q^{\gamma}|\gamma \in \Gamma\} \cup \{q_{ce}^{init}\})$. According to the structure of $BPS$ and $CE$, we know that $Q_{bpns} = ((Q_{b} \cup \{q^{dump}\}) \times \{q^{\gamma}|\gamma \in \Gamma\}) \dot{\cup} ((Q_{b}^{com} \cup \{q^{dump}\}) \times \{q_{ce}^{init}\})$. Thus, we have $|Q_{bpns}| \leq (2^{|Q| \times |Q_{s}|} + 1) \times |\Gamma| + 2^{|Q| \times |Q_{s}|} + 1 = (2^{|Q| \times |Q_{s}|} + 1)(|\Gamma| + 1)$. 
For convenience, we shall call $Q_{bpns}^{rea} := (Q_{b} \cup \{q^{dump}\}) \times \{q^{\gamma}|\gamma \in \Gamma\}$ the set of reaction states since any event, if defined at these states, belongs to $\Sigma$, and $Q_{bpns}^{com} := (Q_{b}^{com} \cup \{q^{dump}\}) \times \{q_{ce}^{init}\}$ the set of control states since any event, if defined at these states, belongs to $\Gamma$. Thus, $Q_{bpns} = Q_{bpns}^{rea} \dot{\cup} Q_{bpns}^{com}$. 

\textbf{Remark III.3:} The above method starts from $P_{\Sigma_{o}}(G||S)$ by adding all the possible control commands that preserve control equivalence. 
An alternative construction of $BPNS$ is to start from $P_{\Sigma_{o} \cup \Gamma}(G||CE)$, which considers all the possible control commands beforehand, 
by pruning the control commands based on comparison with $P_{\Sigma_{o}}(G||S)$ to ensure control equivalence. 
The details are omitted here for brevity.

\begin{figure}[htbp]
\begin{center}
\includegraphics[height=6.5cm]{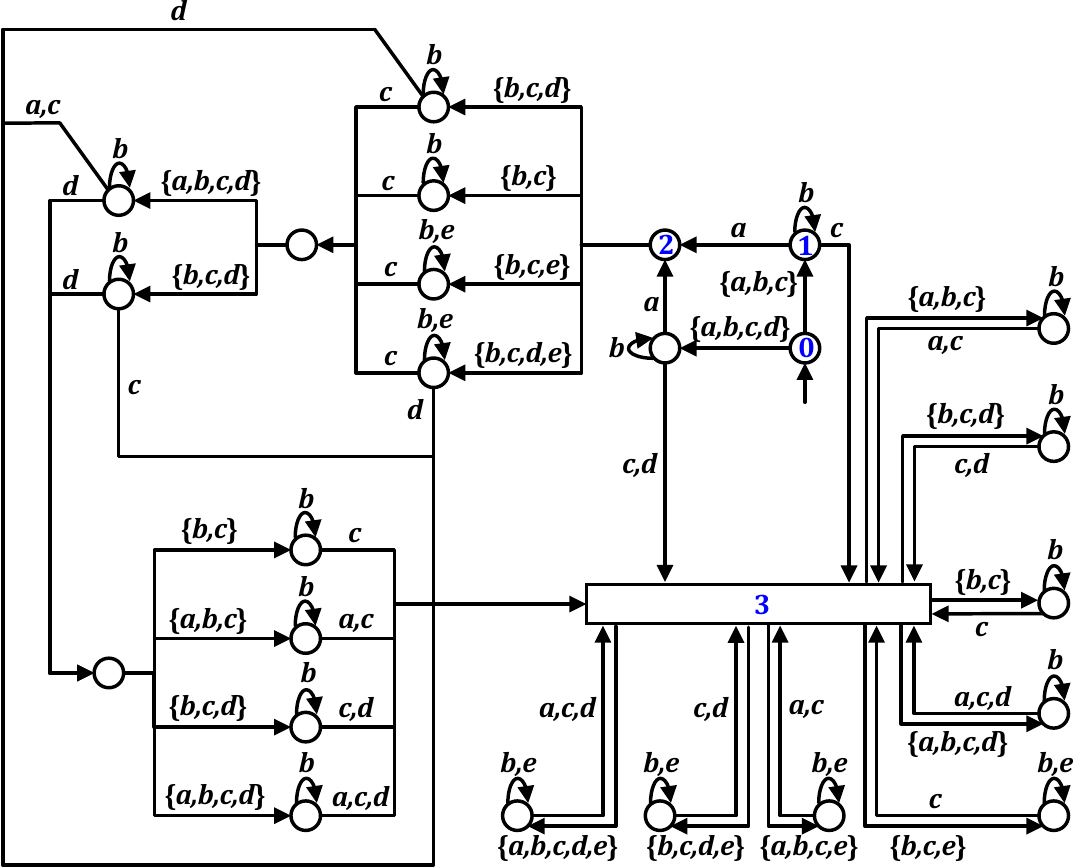}   
\caption{Bipartite behavior-preserving command-nondeterministic supervisor $BPNS$}
\label{fig:BPNS}
\end{center}        
\end{figure}

\textbf{Example III.4} Based on $BPS$ shown in Fig. \ref{fig:BPS} and $CE$ shown in Fig. \ref{fig:CE_CEA}. (a), the computed $BPNS$ is illustrated in Fig. \ref{fig:BPNS}. At the initial control state 0, two control commands $\{a,b,c\}$ and $\{a,b,c,d\}$ are defined, which means that a control equivalent supervisor could issue either $\{a,b,c\}$ or $\{a,b,c,d\}$ when the system initiates. If $\{a,b,c\}$ is issued, then $BPNS$ would transit to state 1, where according to the structure of a bipartite supervisor, unobservable event $b$ in $\{a,b,c\}$ is a self-loop transition, and observable events $a$ and $c$ in $\{a,b,c\}$ lead to two control states, state $2$ and state 3, respectively. At control state 2, to ensure the control equivalence, a supervisor could only issue one of $\{b,c,d\}$, $\{b,c\}$, $\{b,c,e\}$ and $\{b,c,d,e\}$, as encoded in $BPS$. At control state 3, according to the structure of $G$, we know that event $c$ would never occur under the issued initial command $\{a,b,c\}$, thus, any command could be issued without violating the control equivalence. 

In the rest of this subsection, we show $BPNS$ indeed encodes all the control equivalent bipartite supervisors. We have the following results.

\textbf{Lemma III.1:} Given $G$ and $S$, for any supervisor $S'$, we have $L(G||S) = L(G||S')$ iff $L(P_{\Sigma_{o}}(G||S)) = L(P_{\Sigma_{o}}(G||S'))$.

\emph{Proof:} See Appendix \ref{appendix: Lemma III.1}. \hfill $\blacksquare$

\textbf{Proposition III.1:} Given $G$ and $S$, for any supervisor $S' = (Q_{s'}, \Sigma, \xi_{s'}, q_{s'}^{init})$ such that $L(G||S) = L(G||S')$, we have $L(BT(S')) \subseteq L(BPNS)$.


\emph{Proof:} See Appendix \ref{appendix: Proposition III.1}. \hfill $\blacksquare$

\textbf{Proposition III.2:} Given $G$ and $S$, for any supervisor $S' = (Q_{s'}, \Sigma, \xi_{s'}, q_{s'}^{init})$ such that $L(G||S) \neq L(G||S')$, we have $L(BT(S')) \not\subseteq L(BPNS)$.

\emph{Proof:} See Appendix \ref{appendix: Proposition III.2}. \hfill $\blacksquare$ 

In the following text, we shall denote by $\mathscr{S}$ the set of all supervisors that satisfy \
controllability and observability, and $\mathscr{S}_{e}(S) := \{S' \in \mathscr{S} |L(G||S) = L(G||S')\}$ the set of supervisors that are control equivalent to $S$.

\textbf{Theorem III.1:} $\bigcup\limits_{S' \in \mathscr{S}_{e}(S)}L(BT(S')) = L(BPNS)$.

\emph{Proof:} See Appendix \ref{appendix: Theorem III.1}. \hfill $\blacksquare$ 

Based on \textbf{Theorem III.1}, $BPNS$ encodes exactly all the control equivalent bipartite supervisors.


\section{Supervisor Obfuscation Against Covert Actuator Attackers}
\label{sec:Synthesis of Obfuscated Supervisors Against Covert Actuator Attackers}

In the section, based on the bipartite behavior-preserving command-nondeterministic supervisor $BPNS$ constructed in Section \ref{subsec:Generation of control equivalent bipartite supervisors}, we shall firstly find all the damage strings that could be taken use of by  covert damage-reachable actuator attackers for the control equivalent bipartite supervisors encoded in $BPNS$. Then, according to those damage strings, we shall extract from $BPNS$ those control equivalent bipartite supervisors that are resilient against any covert actuator attack.

\subsection{Damage strings encoding}
\label{subsec:Synthesis of covert actuator attackers for control equivalent bipartite supervisors}

We firstly construct the version of $BPNS$ under actuator attack, denoted by $BPNS^{A} = (Q_{bpns}^{a}, \Sigma_{bpns}^{a}, \xi_{bpns}^{a}, q_{bpns}^{a,init})$, where
\begin{enumerate}[1.]
\setlength{\itemsep}{3pt}
\setlength{\parsep}{0pt}
\setlength{\parskip}{0pt}
    \item $Q_{bpns}^{a} = Q_{bpns} \cup \{q_{bpns}^{detect}\} = Q_{bpns}^{rea} \cup Q_{bpns}^{com} \cup \{q_{bpns}^{detect}\}$, where $Q_{bpns}^{rea} = (Q_{b} \cup \{q^{dump}\}) \times \{q^{\gamma}|\gamma \in \Gamma\}$ and $Q_{bpns}^{com} = (Q_{b}^{com} \cup \{q^{dump}\}) \times \{q_{ce}^{init}\}$ 
    \item $\Sigma_{bpns}^{a} = \Sigma \cup \Gamma$
    \item \begin{enumerate}[a.]
        \setlength{\itemsep}{3pt}
        \setlength{\parsep}{0pt}
        \setlength{\parskip}{0pt}
        \item $(\forall q, q' \in Q_{bpns}^{a})(\forall \sigma \in \Sigma \cup \Gamma) \xi_{bpns}(q, \sigma) = q' \Rightarrow \xi_{bpns}^{a}(q, \sigma) = q'$ 
        \item $(\forall q \in Q_{bpns}^{rea})(\forall \sigma \in \Sigma_{c,a} \cap \Sigma_{uo}) \neg\xi_{bpns}(q, \sigma)! \Rightarrow \xi_{bpns}^{a}(q, \sigma) = q$ 
        \item $(\forall q \in Q_{bpns}^{rea})(\forall \sigma \in \Sigma_{o}) \neg\xi_{bpns}(q, \sigma)! \Rightarrow \xi_{bpns}^{a}(q, \sigma) =\\ q_{bpns}^{detect}$
    \end{enumerate}
    \item $q_{bpns}^{a,init} = q_{bpns}^{init}$
\end{enumerate}
The construction procedure of $BPNS^{A}$ from $BPNS$ is similar to that of generating $BT(S)^{A}$ from $BT(S)$ in Section \ref{subsubsec:Supervisor}. Similarly, when $BPNS^{A}$ reaches the state $q_{bpns}^{detect}$, it means that some observation that should not have occurred happens based on the supervisor structure, i.e., the attacker is detected. We note that $(Q_{b} \cup \{q^{dump}\}) \times \{q^{\gamma}|\gamma \in \Gamma\}$ is the set of reaction states and $(Q_{b}^{com} \cup \{q^{dump}\}) \times \{q_{ce}^{init}\}$ is the set of control states in $BPNS^{A}$. We have the following.

\textbf{Proposition IV.1:} Given $G$ and $S$, for any supervisor $S' = (Q_{s'}, \Sigma, \xi_{s'}, q_{s'}^{init})$ such that $L(G||S) = L(G||S')$, we have $L(BT(S')^{A}) \subseteq L(BPNS^{A})$. 

\emph{Proof:} See Appendix \ref{appendix: Proposition IV.1}. \hfill $\blacksquare$

\textbf{Theorem IV.1:} $\bigcup\limits_{S' \in \mathscr{S}_{e}(S)}L(BT(S')^{A}) = L(BPNS^{A})$. 

\emph{Proof:} See Appendix \ref{appendix: Theorem IV.1}. \hfill $\blacksquare$

\begin{figure}[htbp]
\begin{center}
\includegraphics[height=6.5cm]{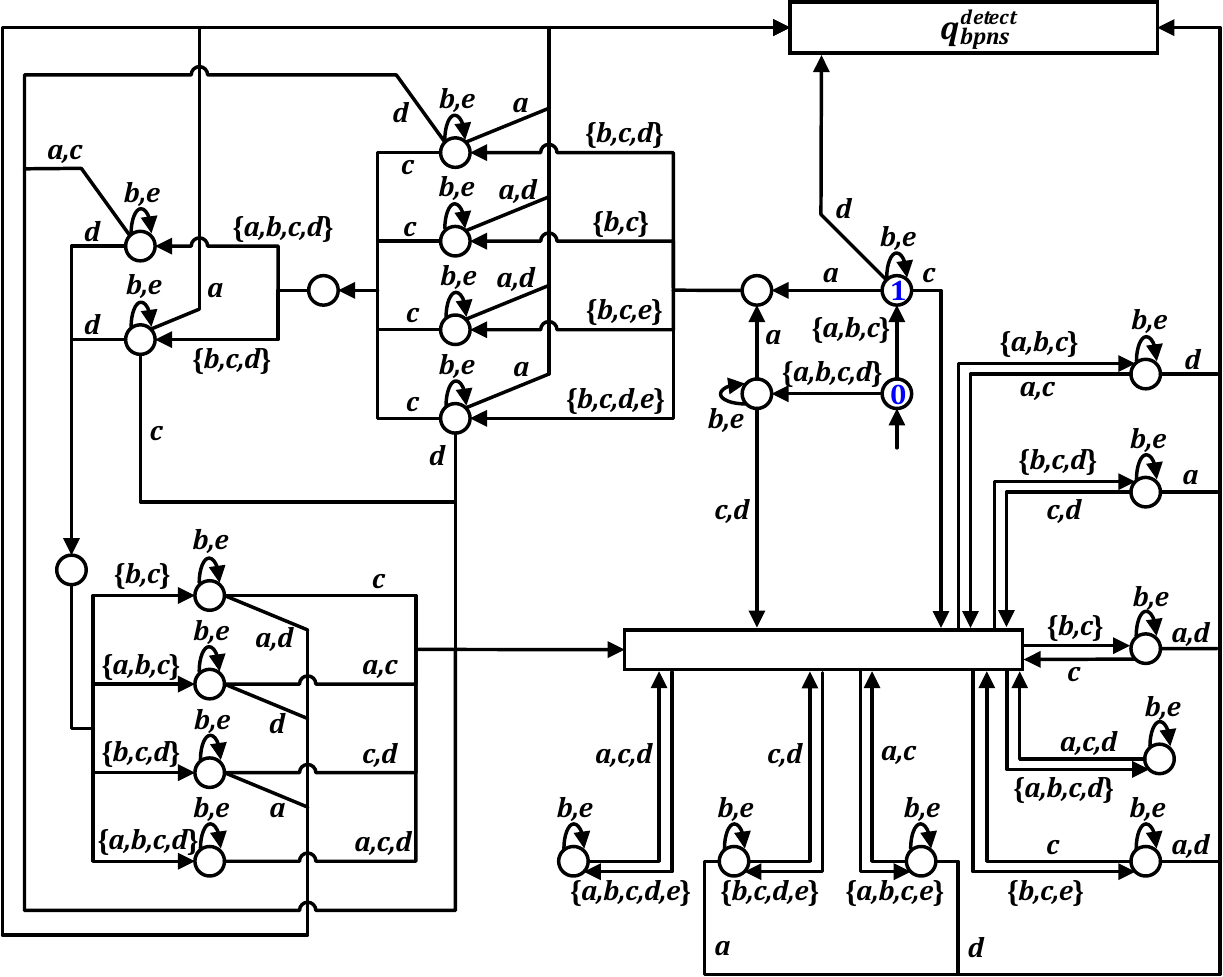}   
\caption{Bipartite behavior-preserving command-nondeterministic supervisor under attack $BPNS^{A}$}
\label{fig:BPNSA}
\end{center}        
\end{figure}

\textbf{Example IV.1} Based on the computed $BPNS$ shown in Fig. \ref{fig:BPNS}, the constructed $BPNS^{A}$ is illustrated in Fig. \ref{fig:BPNSA}. We shall briefly explain the construction procedure by taking state 0 and state 1 as an illustration. According to Case 3.a, 1) at the initial state 0, the control commands $\{a,b,c\}$ and $\{a,b,c,d\}$ originally defined at state 0 in $BPNS$ are retained, 2) at the state 1, the events $a$, $b$ and $c$ originally defined at state 1 are retained. According to Case 3.b, at state 1, the transition labelled by the unobservable but attackable event $e$ is added, which is a self-loop, meaning that $e$ could be enabled by the attacker but would not be observed by the supervisor. According to Case 3.c, the transition labelled by the observable event $d$, which is not originally defined at state 1 in $BPNS$, is added at state 1 and leads to the state $q_{bpns}^{detect}$, meaning that the attacker is discovered.

Next, we shall synthesize a structure to contain all the damage strings that can be taken use of by covert and damage-reachable actuator attackers for  the control equivalent bipartite supervisors encoded in $BPNS$. The procedure is presented as follows:

\noindent \textbf{Procedure 1:}
\begin{enumerate}[1.]
\setlength{\itemsep}{3pt}
\setlength{\parsep}{0pt}
\setlength{\parskip}{0pt}
    \item Compute $\mathcal{P} = G||CE^{A}||BPNS^{A} = (Q_{\mathcal{P}}, \Sigma_{\mathcal{P}}, \xi_{\mathcal{P}}, q_{\mathcal{P}}^{init}, \\Q_{\mathcal{P},m})$, where $Q_{\mathcal{P},m} = Q_{d} \times Q_{ce}^{a} \times Q_{bpns}^{a}$.  
    \item Generate $\mathcal{P}_{r} = (Q_{\mathcal{P}_{r}}, \Sigma_{\mathcal{P}_{r}}, \xi_{\mathcal{P}_{r}}, q_{\mathcal{P}_{r}}^{init}, Q_{\mathcal{P}_{r},m})$.
    \begin{itemize}
    \setlength{\itemsep}{3pt}
    \setlength{\parsep}{0pt}
    \setlength{\parskip}{0pt}
        \item $Q_{\mathcal{P}_{r}} = Q_{\mathcal{P}} - Q_{bad}$
        \begin{itemize}
        \setlength{\itemsep}{3pt}
        \setlength{\parsep}{0pt}
        \setlength{\parskip}{0pt}
            \item $Q_{bad} = \{(q, q_{ce,a}, q_{bpns}^{a}) \in Q_{\mathcal{P}}|q_{bpns}^{a} = q_{bpns}^{detect}\}$
        \end{itemize}
        \item $\Sigma_{\mathcal{P}_{r}} = \Sigma_{\mathcal{P}}$
        \item $(\forall q, q' \in Q_{\mathcal{P}_{r}})(\forall \sigma \in \Sigma_{\mathcal{P}_{r}}) \, \xi_{\mathcal{P}}(q, \sigma) = q' \Leftrightarrow \xi_{\mathcal{P}_{r}}(q, \sigma) = q'$
        \item $q_{\mathcal{P}_{r}}^{init} = q_{\mathcal{P}}^{init}$
        \item $Q_{\mathcal{P}_{r},m} = Q_{\mathcal{P},m} - Q_{bad}$
    \end{itemize}
    \item Synthesize the supremal supervisor ${\hat{\mathcal{A}}} = (Q_{\hat{a}}, \Sigma_{\hat{a}}, \xi_{\hat{a}}, q_{\hat{a}}^{init})$ over the (attacker's) control constraint $\mathscr{C}_{ac}  = (\Sigma_{c,a}, \Sigma_{o,a} \cup \Gamma)$ by treating $\mathcal{P}$ as the plant and $\mathcal{P}_{r}$ as the requirement such that $\mathcal{P}||{\hat{\mathcal{A}}}$ is marker-reachable and safe w.r.t. $\mathcal{P}_{r}$.
\end{enumerate}
We shall briefly explain \textbf{Procedure 1}. At Step 1, we generate a new plant $\mathcal{P} = G||CE^{A}||BPNS^{A}$. At Step 2, we generate $\mathcal{P}_{r}$ from $\mathcal{P}$ by removing those states in $Q_{bad}$, where the covertness is broken, denoted by $q_{bpns}^{a} = q_{bpns}^{detect}$. Then we synthesize the supremal supervisor ${\hat{\mathcal{A}}}$ at Step 3 by treating $\mathcal{P}$ as the plant and $\mathcal{P}_{r}$ as the requirement, whose existence is guaranteed because the set of controllable events $\Sigma_{c,a}$ is a subset of the set of observable events $\Sigma_{o,a} \cup \Gamma$ in the (attacker's) control constraint. Here, we note that, since the target of this work is to find the supervisors that should be resilient against any covert actuator attack, we shall consider the attack in the worst case, i.e., we should find damage strings that could be used by all the covert damage-reachable actuator attackers. Thus, at Step 3, we only need to make sure that $\mathcal{P}||{\hat{\mathcal{A}}}$ is marker-reachable and safe w.r.t. $\mathcal{P}_{r}$.

In the following text, the set of covert and damage-reachable actuator attackers against the supervisor $S'$ (or bipartite supervisor $BT(S')$) w.r.t. the attack constraint $(\Sigma_{o,a}, \Sigma_{c,a})$ is denoted as $\mathscr{A}(S')$ (or $\mathscr{A}(BT(S'))$).

\textbf{Proposition IV.2:} Given $G$ and $S$, for any supervisor $S'$ such that $L(G||S) = L(G||S')$ and any attacker $\mathcal{A} \in \mathscr{A}(S')$, we have $L(G||CE^{A}||BT(S')^{A}||\mathcal{A}) \subseteq L(G||CE^{A}||BPNS^{A}||\hat{\mathcal{A}})$. 

\emph{Proof:} See Appendix \ref{appendix: Proposition IV.2}. \hfill $\blacksquare$

\textbf{Corollary IV.1:} Given $G$ and $S$, for any supervisor $S'$ such that $L(G||S) = L(G||S')$ and any attacker $\mathcal{A} \in \mathscr{A}(S')$, we have $L_{m}(G||CE^{A}||BT(S')^{A}||\mathcal{A}) \subseteq L_{m}(G||CE^{A}||BPNS^{A}||\hat{\mathcal{A}})$.

\emph{Proof:} Based on \textbf{Proposition IV.2}, it holds that LHS $= L(G||CE^{A}||BT(S')^{A}||\mathcal{A}) \cap L_{m}(G) \subseteq L(G||CE^{A}||BPNS^{A}\\||\hat{\mathcal{A}}) \cap L_{m}(G) = L_{m}(G||CE^{A}||BPNS^{A}||\hat{\mathcal{A}}) =$ RHS. \hfill $\blacksquare$

\textbf{Theorem IV.2:} Given $G$ and $S$, we have  
\[
\begin{aligned}
\bigcup\limits_{S' \in \mathscr{S}_{e}(S)}\bigcup\limits_{\mathcal{A} \in \mathscr{A}(S')}&L_{m}(G||CE^{A}||BT(S')^{A}||\mathcal{A}) \\= & L_{m}(G||CE^{A}||BPNS^{A}||\hat{\mathcal{A}})
\end{aligned}
\]

\emph{Proof:} See Appendix \ref{appendix: Theorem IV.2}. \hfill $\blacksquare$

Based on \textbf{Theorem IV.2}, $L_{m}(G||CE^{A}||BPNS^{A}||\hat{\mathcal{A}})$ encodes all the damage strings in all the control equivalent bipartite supervisors, which can be taken use of by the attacker to cause damage infliction. Thus, we can rely on $L_{m}(G||CE^{A}||BPNS^{A}||\hat{\mathcal{A}})$ to remove  inappropriate control commands in $BPNS^{A}$, by which all the control equivalent bipartite supervisors that are resilient against any covert actuator attack can be generated.

\begin{figure}[htbp]
\begin{center}
\includegraphics[height=5.5cm]{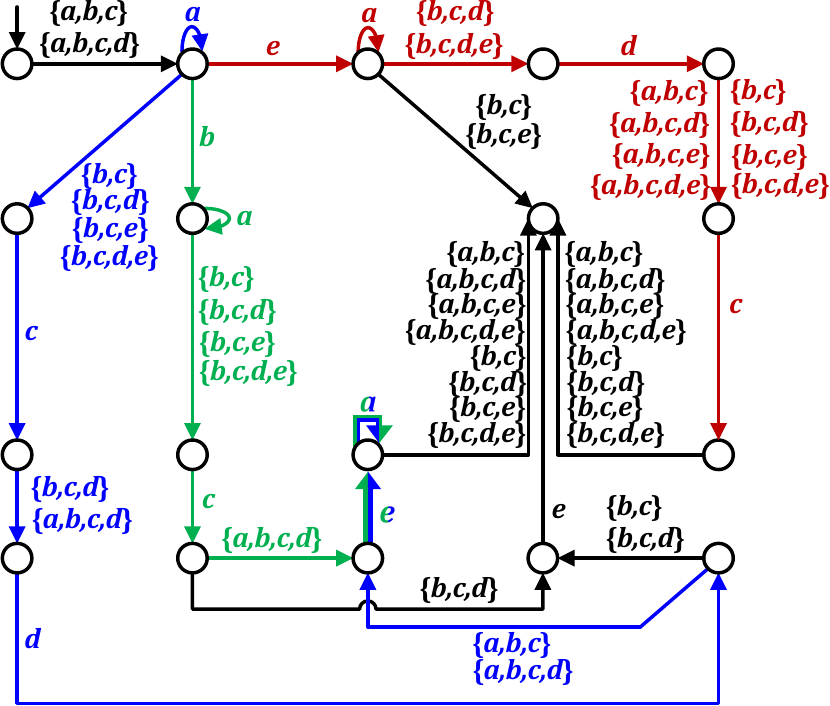}   
\caption{The synthesized $\hat{\mathcal{A}}$}
\label{fig:A}
\end{center}        
\end{figure}

\textbf{Example IV.2} Based on $G$, $CE^{A}$ and $BPNS^{A}$ shown in Fig. \ref{fig:G_S}. (a), Fig. \ref{fig:CE_CEA}. (b) and Fig. \ref{fig:BPNSA}, respectively, the synthesized $\hat{\mathcal{A}}$ by adopting \textbf{Procedure 1} is illustrated in Fig. \ref{fig:A}. It can be checked that $\hat{\mathcal{A}}$ encodes three kinds of damage strings (marked by red, green and blue) that can be taken use of by the attacker: 
\begin{itemize}
\setlength{\itemsep}{3pt}
\setlength{\parsep}{0pt}
\setlength{\parskip}{0pt}
    \item Red part: After observing the initial control command $\{a,b,c\}$ or $\{a,b,c,d\}$, the attacker can carry out the enablement attack to enable the execution of unobservable event $e$, which results in the reuse of initial control command and event $a$ is executed. After that, if the supervisor issues a control command containing event $d$, that is, command $\{b,c,d\}$ or $\{b,c,d,e\}$, then event $d$ would be executed, triggering the command sending by the supervisor and finally event $c$ is executed, causing the damage inflcition. 
    \item Green part: At first, the attacker would not implement any attack, and the system runs the string $\{a,b,c\}/\{a,b,c,d\} \rightarrow b \rightarrow a \rightarrow \{b,c\}/\{b,c,d\}/\{b,c,e\}/\{b,c,d,e\} \rightarrow c$. Afterwards, if the supervisor issues a control command containing event $a$, that is, command $\{a,b,c,d\}$, then the attacker enables the execution of unobservable event $e$, which results in the reuse of command $\{a,b,c,d\}$ and then event $a$ is executed, causing the damage infliction.
    \item Blue part: The idea of this attack strategy is similar to the green part. At first, the attacker would not implement any attack, and the system runs the string $\{a,b,c\}/\{a,b,c,d\} \rightarrow a \rightarrow \{b,c\}/\{b,c,d\}/\{b,c,e\}/\{b,c,d,e\} \rightarrow c \rightarrow \{b,c,d\}/\{a,b,c,d\} \rightarrow d$. Afterwards, if the supervisor issues a control command containing event $a$, that is, command $\{a,b,c\}$ or $\{a,b,c,d\}$, then the attacker enables the execution of unobservable event $e$, which results in the reuse of command $\{a,b,c\}$ or $\{a,b,c,d\}$ and then event $a$ is executed, causing the damage infliction.
\end{itemize}


\subsection{Generation of obfuscated supervisors against covert actuator attackers}
\label{subsec:Generation of Obfuscated Supervisors Against Covert Actuator Attackers}

In this subsection, we shall propose an approach for computing all the resilient supervisors that are control equivalent to $S$. Since $BPNS^{A}$ has exactly encoded all the control equivalent bipartite supervisors under attack based on \textbf{Theorem IV.1}, we know that, with the guidance of $L_{m}(G||CE^{A}||BPNS^{A}||\hat{\mathcal{A}})$, the resilient and control equivalent bipartite supervisors under attack can be obtained from $BPNS^{A}$ by pruning inappropriate transitions that are labelled by commands in $\Gamma$, which are controllable to the supervisor. Thus, the intuitive idea of our methodology is to extract the resilient and control equivalent bipartite supervisors under attack by treating $BPNS^{A}$ as a plant, and then perform the synthesis. 
The detailed methodology is presented as follows. 

\noindent \textbf{Procedure 2:}

\begin{enumerate}[1.]
\setlength{\itemsep}{3pt}
\setlength{\parsep}{0pt}
\setlength{\parskip}{0pt}
    \item Compute $\mathcal{P} = G||CE^{A}||BPNS^{A}||\hat{\mathcal{A}} = (Q_{\mathcal{P}}, \Sigma \cup \Gamma, \xi_{\mathcal{P}}, q_{\mathcal{P}}^{init}, Q_{\mathcal{P},m})$.
    \item Construct $\mathcal{P}_{r} = (Q_{\mathcal{P}_{r}}, \Sigma \cup \Gamma, \xi_{\mathcal{P}_{r}}, q_{\mathcal{P}_{r}}^{init})$ based on $\mathcal{P}$, where 
    \begin{enumerate}[a.]
    \setlength{\itemsep}{3pt}
    \setlength{\parsep}{0pt}
    \setlength{\parskip}{0pt}
        \item $Q_{\mathcal{P}_{r}} = (Q_{\mathcal{P}} - Q_{\mathcal{P},m}) \cup \{q^{dump}\}$
        \item \begin{enumerate}[i.]
        \setlength{\itemsep}{3pt}
        \setlength{\parsep}{0pt}
        \setlength{\parskip}{0pt}
            \item $(\forall q, q' \in Q_{\mathcal{P}} - Q_{\mathcal{P},m})(\forall \sigma \in \Sigma \cup \Gamma)\xi_{\mathcal{P}}(q, \sigma) = q' \Rightarrow \xi_{\mathcal{P}_{r}}(q, \sigma) = q'$
            \item $(\forall q \in Q_{\mathcal{P}} - Q_{\mathcal{P},m})(\forall \sigma \in \Sigma \cup \Gamma)\neg\xi_{\mathcal{P}}(q, \sigma)! \Rightarrow \xi_{\mathcal{P}_{r}}(q, \sigma) = q^{dump}$ 
            \item $(\forall \sigma \in \Sigma \cup \Gamma)\xi_{\mathcal{P}_{r}}(q^{dump}, \sigma) = q^{dump}$
        \end{enumerate}
        \item $q_{\mathcal{P}_{r}}^{init} = q_{\mathcal{P}}^{init}$
    \end{enumerate}
    \item Synthesize the supremal supervisor $S_{0}^{A} = (Q_{S_{0}^{A}}, \Sigma \cup \Gamma, \xi_{S_{0}^{A}}, q_{S_{0}^{A}}^{init})$ over the control constraint $(\Gamma, \Sigma_{o} \cup \Gamma)$ by treating $BPNS^{A}$ as the plant and $\mathcal{P}_{r}$ as the requirement such that $BPNS^{A}||S_{0}^{A}$ is safe w.r.t. $\mathcal{P}_{r}$.
\end{enumerate}  
In this procedure, we shall prune  control commands in $BPNS^{A}$ to obtain all the  resilient and control equivalent  bipartite supervisors under attack encoded in $BPNS^{A}$. At Step 1, we compute $\mathcal{P} = G||CE^{A}||BPNS^{A}||\hat{\mathcal{A}}$, whose marked behavior encodes all the damage strings that could lead to damage infliction for  control equivalent supervisors. Next, to find all the resilient and control equivalent bipartite supervisors under attack encoded in $BPNS^{A}$, we shall ``design a supervisor'' to ``control'' $BPNS^{A}$, which encodes all the control equivalent bipartite supervisors under attack, by disabling control commands such that the damage strings would not occur in the plant $BPNS^{A}$. Thus, at Step 2, we construct a requirement automaton $\mathcal{P}_{r}$ based on $\mathcal{P}$, where 1) the set of marker states of $\mathcal{P}$ are removed in $\mathcal{P}_{r}$ in Step 2.a, 2) all the transitions originally defined in $\mathcal{P}$ are retained in $\mathcal{P}_{r}$ for those states that have not been removed, as shown in Step 2.b.i, 3) for those states that have not been removed, we complete the transitions that are not originally defined in $\mathcal{P}$, which would lead to the newly added state $q^{dump}$, as shown in Step 2.b.ii, and 4) all the transitions in $\Sigma \cup \Gamma$ are defined at the state $q^{dump}$ in Step 2.b.iii. 
The purpose of adding Step 2.b.ii and Step 2.b.iii to complete transitions is: The requirement in the synthesis is only supposed to forbid the execution of the strings that might result in damage infliction. Thus, we carry out Step 2.b.ii and Step 2.b.iii such that $\mathcal{P}_{r}$ specifies the set of all the possible legal strings. 
At Step 3, by treating $BPNS^{A}$ as the plant and $\mathcal{P}_{r}$ as the requirement, we could synthesize a safe supervisor $S_{0}^{A}$. Since the requirement automaton has removed all the damage strings, $S_{0}^{A}$ indeed includes 
the attacked version of those resilient and control equivalent bipartite supervisors in $BPNS$.

\textbf{Remark IV.1:} Since $S_{0}^{A}$ is synthesized by treating $BPNS^{A}$ as the plant at Step 3 of \textbf{Procedure 2}, we consider the case where the synthesized supervisor $S_{0}^{A}$ satisfies the requirement that $L(S_{0}^{A}) \subseteq L(BPNS^{A})$,  following the standard notion of controllability and observability \cite{WMW10} over the control constraint $(\Gamma, \Sigma_{o} \cup \Gamma)$ w.r.t. the plant $BPNS^{A}$. Without loss of generality, any event in $\Sigma_{uo}$, if defined, is a self-loop transition in $S_{0}^{A}$. Thus, $S_{0}^{A}$ is a bipartite structure\footnote{If we follow our definition of a supervisor and synthesize $S_{0}^{A}$, we could always update $S_{0}^{A} := BPNS^{A}||S_{0}^{A}$ to generate a bipartite supervisor $S_{0}^{A}$ with $L(S_{0}^{A}) \subseteq L(BPNS^{A})$.} similar to $BPNS^{A}$.

\textbf{Example IV.3} Based on $G$, $CE^{A}$, $BPNS^{A}$ and $\hat{\mathcal{A}}$ shown in Fig. \ref{fig:G_S}. (a), Fig. \ref{fig:CE_CEA}. (b), Fig. \ref{fig:BPNSA} and Fig. \ref{fig:A}, respectively, the synthesized $S_{0}^{A}$ by adopting \textbf{Procedure 2} is illustrated in Fig. \ref{fig:S0A}. Compared with $BPNS^{A}$ shown in Fig. \ref{fig:BPNSA}, to ensure the damage infliction would not be caused, there are several control commands removed at some control states in $S_{0}^{A}$:
\begin{figure}[htbp]
\begin{center}
\includegraphics[height=5.5cm]{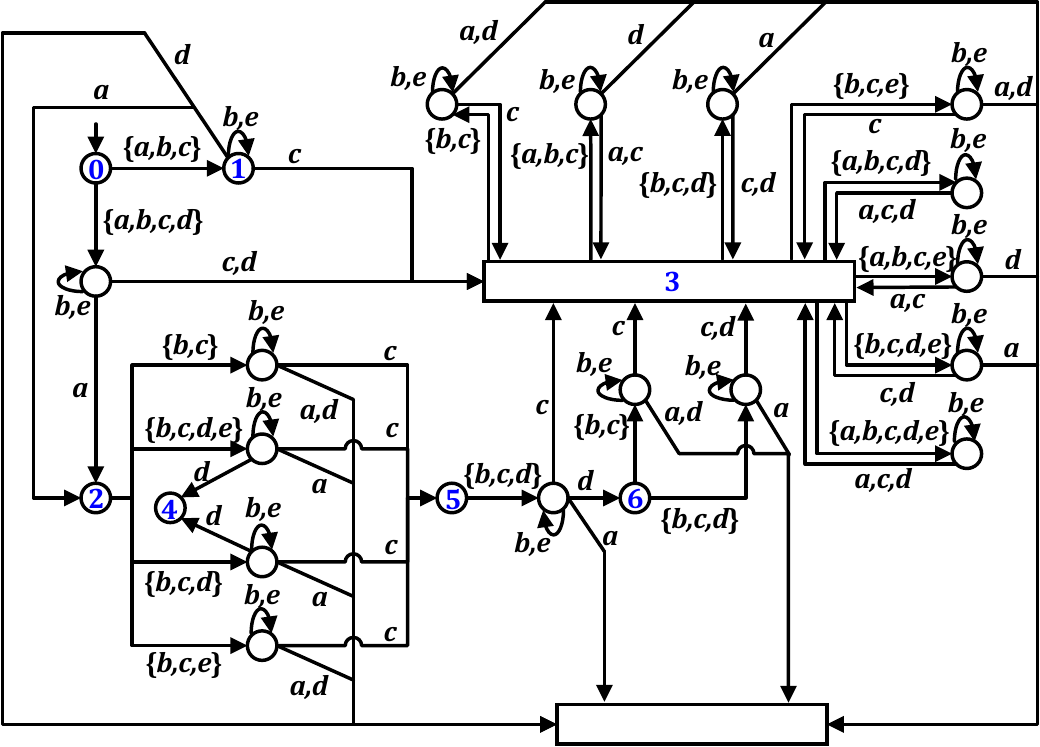}   
\caption{The synthesized $S_{0}^{A}$}
\label{fig:S0A}
\end{center}        
\end{figure}
\begin{itemize}
\setlength{\itemsep}{3pt}
\setlength{\parsep}{0pt}
\setlength{\parskip}{0pt}
    \item After $S_{0}^{A}$ observes the sequence $\{a,b,c\}/\{a,b,c,d\} \rightarrow a \rightarrow \{b,c,d\}/\{b,c,d,e\} \rightarrow d$, $S_{0}^{A}$ transits to state 4, where now the plant might execute the string $ead$ due to the enablement attack of unobservable event $e$. Thus, $S_{0}^{A}$ cannot issue any control command at state 4 as the string $eadc$ would cause damage infliction and uncontrollable event $c$ is contained in any control command. This case corresponds to the defense strategy against the damage strings marked by red explained in \textbf{Example IV.2}.
    \item After $S_{0}^{A}$ observes the sequence $s = \{a,b,c\}/\{a,b,c,d\} \rightarrow a \rightarrow \{b,c\}/\{b,c,d\}/\{b,c,e\}/\{b,c,d,e\} \rightarrow c$, $S_{0}^{A}$ transits to state 5, where now the plant might execute the string $bac$ as $b$ is unobservable to the supervisor. Since the string $bacea$ would cause damage infliction and unobservable event $e$ could be enabled by the attacker, $S_{0}^{A}$ cannot issue any control command containing event $a$ at state 5, that is, the command $\{a,b,c,d\}$, which follows the sequence $s$ in $BPNS^{A}$, cannot be defined at state 5 and only command $\{b,c,d\}$ is retained. This case corresponds to the defense strategy against the damage strings marked by green explained in \textbf{Example IV.2}.
    \item After $S_{0}^{A}$ observes the sequence $s = \{a,b,c\}/\{a,b,c,d\} \rightarrow a \rightarrow \{b,c\}/\{b,c,d\}/\{b,c,e\}/\{b,c,d,e\} \rightarrow c \rightarrow \{b,c,d\} \rightarrow d$, $S_{0}^{A}$ transits to state 6, where now the plant executes the string $acd$. Since the string $acdea$ would cause damage infliction and unobservable event $e$ could be enabled by the attacker, $S_{0}^{A}$ cannot issue any control command containing event $a$ at state 6, that is, commands $\{a,b,c\}$ and $\{a,b,c,d\}$, which follow the sequence $s$ in $BPNS^{A}$, cannot be defined at state 6 and only commands $\{b,c\}$ and $\{b,c,d\}$ are retained. This case corresponds to the defense strategy against the damage strings marked by blue explained in \textbf{Example IV.2}.
\end{itemize}


Next, we shall transform $S_{0}^{A}$ to the version in the absence of attack. The generated automaton is denoted as $S_{0} = (Q_{S_{0}}, \Sigma \cup \Gamma, \xi_{S_{0}}, q_{S_{0}}^{init})$, where 
\begin{enumerate}[1.]
\setlength{\itemsep}{3pt}
\setlength{\parsep}{0pt}
\setlength{\parskip}{0pt}
    \item $Q_{S_{0}} = Q_{S_{0}^{A}}$
    \item \begin{enumerate}[a.]
    \setlength{\itemsep}{3pt}
    \setlength{\parsep}{0pt}
    \setlength{\parskip}{0pt}
        \item $(\forall q, q' \in Q_{S_{0}})(\forall \gamma \in \Gamma)\xi_{S_{0}^{A}}(q, \gamma) = q' \Rightarrow \xi_{S_{0}}(q, \gamma) = q'$
        \item $(\forall q, q' \in Q_{S_{0}})(\forall \gamma \in \Gamma)(\forall \sigma \in \gamma \cap \Sigma_{uo})\xi_{S_{0}^{A}}(q, \gamma) = q' \Rightarrow \xi_{S_{0}}(q', \sigma) = q'$
        \item $(\forall q, q', q'' \in Q_{S_{0}})(\forall \gamma \in \Gamma)(\forall \sigma \in \gamma \cap \Sigma_{o})\xi_{S_{0}^{A}}(q, \gamma) \\= q' \wedge \xi_{S_{0}^{A}}(q', \sigma) = q'' \Rightarrow \xi_{S_{0}}(q', \sigma) = q''$
    \end{enumerate}
    \item $q_{S_{0}}^{init} = q_{S_{0}^{A}}^{init}$
\end{enumerate}
Briefly speaking, 1) we retain all the transitions labelled by events in $\Gamma$ that are originally defined in $S_{0}^{A}$, as shown in Step 2.a, 2) for any state $q'$ such that there exists a transition $\xi_{S_{0}^{A}}(q, \gamma) = q'$, we retain the transition labelled by any event in $\gamma \cap \Sigma_{uo}$ ($\gamma \cap \Sigma_{o}$, respectively), which is a self-loop (leads to a new state $q''$, respectively), as shown in Step 2.b (Step 2.c, respectively). 
Then we generate the automaton $Ac(S_{0})$. For convenience, in the rest, we shall refer to $Ac(S_{0})$ whenever we talk about $S_{0}$.
By Remark IV.1, $S_{0}$ is a bipartite structure and the state set of $S_{0}$ could be divided into two disjoint sets $Q_{S_{0}} = Q_{S_{0}}^{rea} \dot{\cup} Q_{S_{0}}^{com}$, where $Q_{S_{0}}^{rea}$ is the set of reaction states and $Q_{S_{0}}^{com}$ is the set of control states, and 
\begin{enumerate}[1.]
\setlength{\itemsep}{3pt}
\setlength{\parsep}{0pt}
\setlength{\parskip}{0pt}
    \item At any state of $Q_{S_{0}}^{rea}$, any event in $\Gamma$ is not defined.
    \item At any state of $Q_{S_{0}}^{rea}$, any event in $\Sigma_{uo}$, if defined, leads to a self-loop, and any event in $\Sigma_{o}$, if defined, would lead to a transition to a control state.
    \item At any state of $Q_{S_{0}}^{com}$, any event in $\Sigma$ is not defined.
    \item At any state of $Q_{S_{0}}^{com}$, any event in $\Gamma$, if defined, would lead to a transition to a reaction state.
\end{enumerate}
We shall briefly explain why these 4 facts hold. We know that: 1) since $L(BPNS^{A}) \subseteq \overline{(\Gamma\Sigma_{uo}^{*}\Sigma_{o})^{*}}$, we have $L(S_{0}) \subseteq \overline{(\Gamma\Sigma_{uo}^{*}\Sigma_{o})^{*}}$, 2) any transition labelled by an unobservable event in $\Sigma_{uo}$ would be a self-loop in $S_{0}$ and any transition labelled by an event in $\Sigma_{o} \cup \Gamma$ would enable $S_{0}$ to make a state transition. Thus, we could always divide the state set of $S_{0}$ into two disjoint parts: 1) the set of control states $Q_{S_{0}}^{com}$, where any event in $\Sigma$ is not defined (fact 3) and any event defined at such a state belongs to $\Gamma$, 2) the set of reaction states $Q_{S_{0}}^{rea}$, where any event in $\Gamma$ is not defined (fact 1) and any event defined at such a state belongs to $\Sigma$. In addition, based on the form of closed-behavior of $BPNS^{A}$, fact 2 and fact 4 naturally hold.

\begin{figure}[htbp]
\begin{center}
\includegraphics[height=4.8cm]{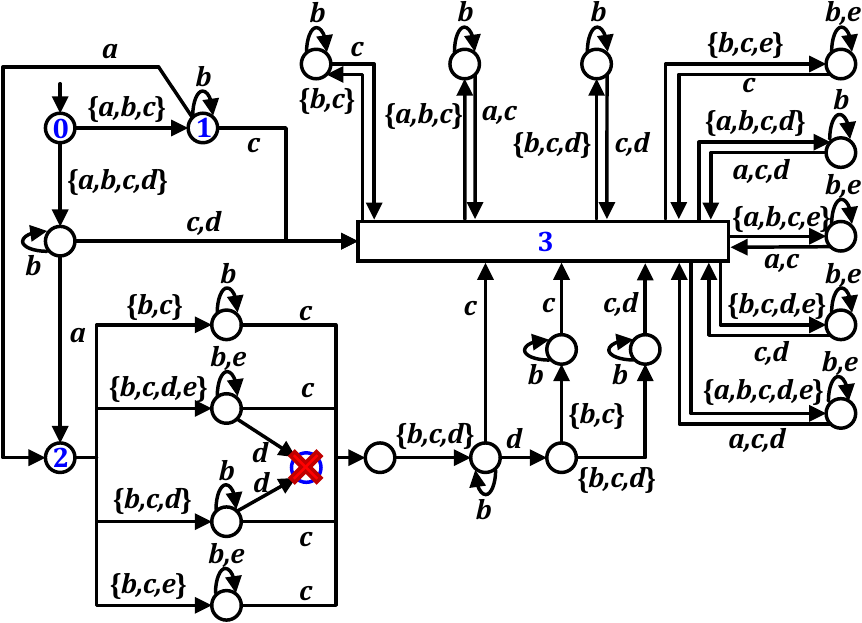}   
\caption{The transformed $S_{0}$}
\label{fig:S0}
\end{center}        
\end{figure}

\textbf{Example IV.4} Based on $S_{0}^{A}$ shown in Fig. \ref{fig:S0A}, the transformed $S_{0}$ is illustrated in Fig. \ref{fig:S0}. By taking several states as an illustration, we shall briefly explain how to obtain $S_{0}$ based on $S_{0}^{A}$. At the initial state 0 of $S_{0}$, according to Case 2.a of the construction of $S_{0}$, we retain the transitions labelled by commands $\{a,b,c\}$ and $\{a,b,c,d\}$ originally defined in $S_{0}^{A}$. After the command $\{a,b,c\}$ is issued, $S_{0}$ would transit to state 1, where we shall only retain those transitions in the absence of attack. According to Case 2.b, the transition labelled by event $b \in \{a,b,c\} \cap \Sigma_{uo}$ is retained, which is a self-loop. According to Case 2.c, the transitions labelled by events $a,c \in \{a,b,c\} \cap \Sigma_{o}$ are retained, which would lead to state 2 and state 3, respectively, as defined in $S_{0}^{A}$.

Although inappropriate control commands that would result in damage infliction have been removed in $S_{0}$, we cannot ensure $S_{0}$ exactly encodes all the resilient and control equivalent bipartite supervisors. This is because it is possible that at some control state of $S_{0}$, where an observation has just been received, there are no control commands defined as a result of the synthesis. Such a phenomenon violates the structure of a bipartite supervisor, where a control command must be defined at any control state according to the construction of a bipartite supervisor as presented in Section \ref{subsubsec:Supervisor}. Thus, by treating $S_{0}$ as a plant, we shall carry out the following procedure to iteratively remove these newly created bad states until the generated structure satisfies the condition that any observation is followed by at least a control command.

\noindent \textbf{Procedure 3:}

\begin{enumerate}[1.]
\setlength{\itemsep}{3pt}
\setlength{\parsep}{0pt}
\setlength{\parskip}{0pt}
    \item Let $k := 0$.
    \item Compute $Q_{k,del} := \{q \in Q_{S_{k}}^{com}|En_{S_{k}}(q) = \varnothing\}$. 
    \item If $Q_{k,del} = \varnothing$, then output $S_{k}$ and end the procedure; otherwise, i.e., $Q_{k,del} \neq \varnothing$, then proceed to Step 4. 
    \item Construct $S_{k,r} = (Q_{S_{k,r}}, \Sigma \cup \Gamma, \xi_{S_{k,r}}, q_{S_{k,r}}^{init})$ based on $S_{k}$, where 
    \begin{enumerate}[a.]
    \setlength{\itemsep}{3pt}
    \setlength{\parsep}{0pt}
    \setlength{\parskip}{0pt}
        \item $Q_{S_{k,r}} = Q_{S_{k}} - Q_{k,del}$
        \item $(\forall q, q' \in Q_{S_{k,r}})(\forall \sigma \in \Sigma \cup \Gamma)\xi_{S_{k}}(q, \sigma) = q' \Rightarrow \xi_{S_{k,r}}(q, \sigma) = q'$
        \item $q_{S_{k,r}}^{init} = q_{S_{k}}^{init}$
    \end{enumerate}
    \item Synthesize the supremal supervisor $S_{k+1} = (Q_{S_{k+1}}, \Sigma \cup\\ \Gamma, \xi_{S_{k+1}}, q_{S_{k+1}}^{init})$ over the control constraint $(\Gamma, \Sigma_{o} \cup \Gamma)$ by treating $S_{k}$ as the plant and $S_{k,r}$ as the requirement such that $S_{k}||S_{k+1}$ is safe w.r.t. $S_{k,r}$. We denote
    $Q_{S_{k+1}} = Q_{S_{k+1}}^{rea} \dot{\cup} Q_{S_{k+1}}^{com}$, where $Q_{S_{k+1}}^{rea}$ is the set of reaction states and $Q_{S_{k+1}}^{com}$ is the set of control states\footnote{The division rule is the same as that of $Q_{S_{0}} = Q_{S_{0}}^{rea} \dot{\cup} Q_{S_{0}}^{com}$.}. 
    \item Let $k \leftarrow k+1$ and proceed to Step 2.
\end{enumerate}
At Step 1, we set the counter $k$ to be 0. Then, at Step 2, taking the $k$-th iteration as an illustration, we shall compute the set of control states in $S_{k}$, denoted by $Q_{k,del}$, where any control state $q \in Q_{k,del}$ satisfies that there are no control commands defined at state $q$, denoted by $En_{S_{k}}(q) = \varnothing$. According to the structure of a bipartite supervisor, any state in $Q_{k,del}$ should not exist in $S_{k}$. At Step 3, if $Q_{k,del} = \varnothing$, then $S_{k}$ is the desired structure that exactly encodes all the resilient and control equivalent bipartite supervisors; otherwise, we shall remove the set of states $Q_{k,del}$ in $S_{k}$ to construct a requirement automaton $S_{k,r}$ for the plant $S_{k}$ at Step 4. At Step 5, by treating $S_{k}$ as the plant and $S_{k,r}$ as the requirement, we shall synthesize a safe supervisor $S_{k+1}$. 

\textbf{Remark IV.2:} Similar to Remark IV.1, we consider the case where the synthesized supervisor $S_{k+1}$ satisfies the requirement that  $L(S_{k+1}) \subseteq L(S_{k})$, following the standard notion of controllability and observability \cite{WMW10} over the control constraint $(\Gamma, \Sigma_{o} \cup \Gamma)$. Without loss of generality, any event in $\Sigma_{uo}$, if defined, is a self-loop transition in $S_{k+1}$. Thus, $S_{k+1}$ is a bipartite structure.


We shall name the output of \textbf{Procedure 3} as obfuscated command non-deterministic supervisor, and we denote it by 
$ONS = (Q_{ons}, \Sigma \cup \Gamma, \xi_{ons}, q_{ons}^{init})$. For convenience, we denote $Q_{ons} = Q_{ons}^{rea} \dot{\cup} Q_{ons}^{com}$, where $Q_{ons}^{rea}$ is the set of reaction states and $Q_{ons}^{com}$ is the set of control states\footnote{The division rule is the same as that of $Q_{S_{0}} = Q_{S_{0}}^{rea} \dot{\cup} Q_{S_{0}}^{com}$.}. 

\textbf{Proposition IV.3:} $L(ONS) \subseteq L(BPNS)$.

\emph{Proof:} See Appendix \ref{appendix: Proposition IV.3}. \hfill $\blacksquare$

\textbf{Theorem IV.3:} $\bigcup\limits_{S' \in \mathscr{S}_{e}^{r}(S)}L(BT(S')) = L(ONS)$, where $\mathscr{S}_{e}^{r}(S)$ denotes the set of resilient supervisors that are control equivalent to $S$.

\emph{Proof:} See Appendix \ref{appendix: Theorem IV.3}. \hfill $\blacksquare$

\textbf{Theorem IV.4:} \textbf{Problem 1} is decidable.

\emph{Proof:} To prove this result, based on \textbf{Theorem IV.3}, we only need to additionally check whether \textbf{Procedure 2} and \textbf{Procedure 3} could terminate within finite steps. Clearly, \textbf{Procedure 2}  terminates within finite steps. For \textbf{Procedure 3}, in each iteration, since the requirement $S_{k,r}$ is generated by removing at least one control state $q$ from the plant $S_{k}$ and any unobservable event in $\Sigma_{uo}$, if defined, is a self-loop in $S_{k}$, we know that $S_{k+1}$ is a substructure of $S_{k}$, and to satisfy the controllability w.r.t. the control constraint $(\Gamma, \Sigma_{o} \cup \Gamma)$, at least two states of $S_{k}$ are removed to get $S_{k+1}$, including the removed control state $q$ and the reaction state $q'$ where there exists $\sigma \in \Sigma_{o}$ such that $\xi_{S_{k}}(q', \sigma) = q$. Thus, \textbf{Procedure 3} would iterate Steps 2-6 for at most $\lfloor \frac{|Q_{S_{0}}|}{2} \rfloor$ times\footnote{$\lfloor \cdot \rfloor$ is the floor function that takes as input a real number, and gives as output the greatest integer less than or equal to this real number.}. This completes the proof. \hfill $\blacksquare$


Next, we shall analyze the computational complexity of the proposed algorithm to synthesize all the obfuscated supervisors, which depends on the complexity of three synthesis steps (\textbf{Procedure 1}, \textbf{Procedure 2} and \textbf{Procedure 3}), and the construction of $S_{0}$ from $S_{0}^{A}$. By using the synthesis approach in \cite{WMW10,WLLW18}, the complexity of \textbf{Procedure 1} is $O((|\Sigma| + |\Gamma|)2^{|Q| \times |Q_{ce}^{a}| \times |Q_{bpns}^{a}|})$, the complexity of \textbf{Procedure 2} is $O((|\Sigma| + |\Gamma|)|Q_{S_{0}^{A}}|)$,
and the complexity of \textbf{Procedure 3} is no more than $O((|\Sigma| + |\Gamma|)|Q_{S_{0}}| + (|\Sigma| + |\Gamma|)(|Q_{S_{0}}|-2) + \dots + (|\Sigma| + |\Gamma|) \times 3)= O((|\Sigma| + |\Gamma|)|Q_{S_{0}}|^{2})$  when $|Q_{S_{0}}|$ is odd 
and no more than $O((|\Sigma| + |\Gamma|)|Q_{S_{0}}| + (|\Sigma| + |\Gamma|)(|Q_{S_{0}}|-2) + \dots + (|\Sigma| + |\Gamma|) \times 2) = O((|\Sigma| + |\Gamma|)|Q_{S_{0}}|^{2})$ when $|Q_{S_{0}}|$ is even. The complexity of constructing $S_{0}$ from $S_{0}^{A}$ is $O((|\Sigma| + |\Gamma|)|Q_{S_{0}^{A}}|)$. Thus, the overall complexity is $O((|\Sigma| + |\Gamma|)|Q_{S_{0}}|^{2})$, where 
\begin{itemize}
\setlength{\itemsep}{3pt}
\setlength{\parsep}{0pt}
\setlength{\parskip}{0pt}
    \item $|Q_{ce}^{a}| = |\Gamma| + 1$
    \item $|Q_{bpns}^{a}| \leq  (2^{|Q| \times |Q_{s}|} + 1)(|\Gamma| + 1) + 1$
   \item $|Q_{S_{0}}| \leq |Q_{S_{0}^{A}}| - 1$
    \item $|Q_{S_{0}^{A}}| \leq 2^{|Q| \times |Q_{ce}^{a}| \times |Q_{bpns}^{a}|^{2} \times |Q_{\hat{a}}|}$ 
    \item $|Q_{\hat{a}}| \leq 2^{|Q| \times |Q_{ce}^{a}| \times |Q_{bpns}^{a}|}$
\end{itemize}

\textbf{Remark IV.3}:  In this work, we focus on addressing the decidability of synthesizing obfuscated supervisors against covert actuator attacks. We are not sure if the above analysis of complexity upper bound is tight. This issue will not be addressed here and is left as a future work.

\begin{figure}[htbp]
\begin{center}
\includegraphics[height=4.2cm]{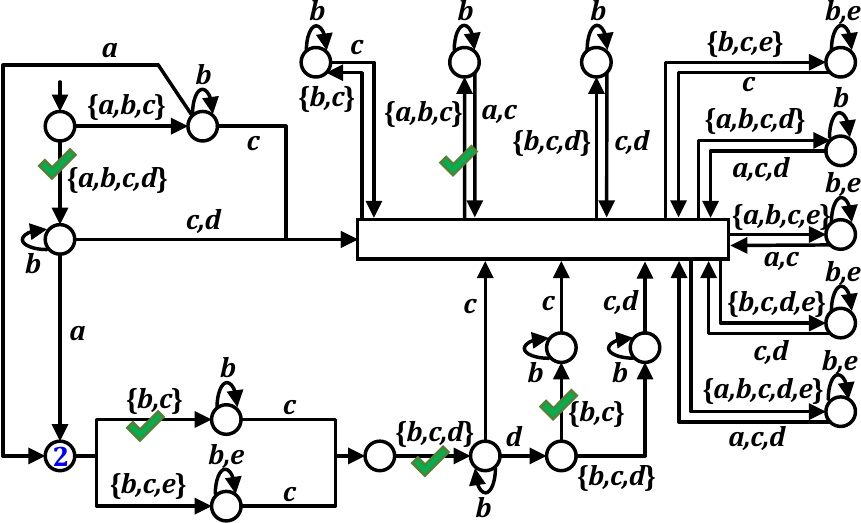}   
\caption{The synthesized $S_{1}$ ($ONS$)}
\label{fig:S1}
\end{center}        
\end{figure}

\textbf{Example IV.5} We shall continue with $S_{0}$ shown in Fig. \ref{fig:S0}. It can be checked that there exists a control state marked by a red cross in $S_{0}$, where there is no control command defined, and this violates the structure of a bipartite supervisor. Thus, according to Step 2 of \textbf{Procedure 3}, this state is included in $Q_{0,del}$ and according to Step 4 of \textbf{Procedure 3}, we remove this state to generate $S_{0,r}$. Then, by treating $S_{0}$ as the plant and $S_{0,r}$ as the requirement, we could synthesize $S_{1}$, which is illustrated in Fig. \ref{fig:S1}. It can be checked that at least a control command is defined at any control state of $S_{1}$, which means that $Q_{1,del} = \varnothing$. Thus, the procedure terminates after the first iteration and outputs $ONS := S_{1}$. Compared with $S_{0}$ shown in Fig. \ref{fig:S0}, the control commands $\{b,c,d\}$ and $\{b,c,d,e\}$ are removed at state 2. Intuitively speaking, this is because 1) at state 2, we know that the supervisor has issued the initial control command $\{a,b,c\}$ or $\{a,b,c,d\}$ and the plant might execute the string $ea$ due to the enablement of unobservable event $e$ by the attacker, 2) if the supervisor issues any command containing the event $d$ at state 2, then $d$ might be executed at $G$ and when the event $d$ is observed by the supervisor, no matter which command is issued by the supervisor, $G$ would execute the event $c$ and reach the damage state as $c$ is uncontrollable and always contained in any control command.

Based on \textbf{Theorem IV.3}, $ONS$ has already exactly encoded all the control equivalent and resilient bipartite supervisors. Next, we show how to extract a control equivalent and resilient bipartite supervisor from $ONS$. We construct the following structure, denoted by $OS = (Q_{os}, \Sigma \cup \Gamma, \xi_{os}, q_{os}^{init})$, where 
\begin{enumerate}[1.]
\setlength{\itemsep}{3pt}
\setlength{\parsep}{0pt}
\setlength{\parskip}{0pt}
    \item $Q_{os} = Q_{ons}$
    \item \begin{enumerate}[a.]
    \setlength{\itemsep}{3pt}
    \setlength{\parsep}{0pt}
    \setlength{\parskip}{0pt}
        \item $(\forall q, q' \in Q_{os})(\forall \sigma \in \Sigma)\xi_{ons}(q, \sigma) = q' \Rightarrow \xi_{os}(q, \sigma) = q'$
        \item For any control state $q \in Q_{ons}^{com}$, we randomly pick a control command $\gamma \in En_{ONS}(q)$ and define that: for any reaction state $q' \in Q_{ons}^{rea}$, if $\xi_{ons}(q, \gamma) = q'$, then $\xi_{os}(q, \gamma) = q'$ and for any control command $\gamma' \in En_{ONS}(q) - \{\gamma\}$, we have $\neg\xi_{os}(q, \gamma')!$.
    \end{enumerate}
    \item $q_{os}^{init} = q_{ons}^{init}$
\end{enumerate}
Then we generate the automaton $Ac(OS)$. For convenience, we shall still denote $Ac(OS)$ as $OS$. The basic idea for the construction of $OS$ from $ONS$ is: at any control state $q$ of $OS$, we shall only retain one transition labelled by a control command originally defined at the state $q$ in $ONS$, and for any other control command $\gamma' \in En_{ONS}(q) - \{\gamma\}$, we do not define $\gamma'$ at the state $q$ in $OS$, as shown in Step 2.b.

\textbf{Proposition IV.4:} Given $G$ and $S$, we have $OS \in \mathscr{S}_{e}^{r}(S)$.

\emph{Proof:} See Appendix \ref{appendix: Proposition IV.4}. \hfill $\blacksquare$

\textbf{Theorem IV.5:} \textbf{Problem 2} is decidable.

\emph{Proof:} Based on \textbf{Theorem IV.4} and \textbf{Proposition IV.4}, we could directly have this result. \hfill $\blacksquare$

\begin{figure}[htbp]
\begin{center}
\includegraphics[height=2.5cm]{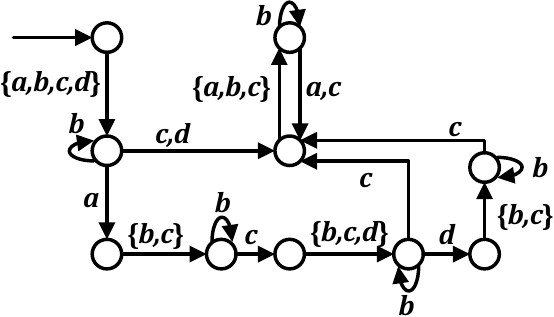}   
\caption{A control equivalent and resilient supervisor $OS$ extracted from $ONS$}
\label{fig:OS}
\end{center}        
\end{figure}

\textbf{Example IV.6} Based on $ONS$ shown in Fig. \ref{fig:S1}, by choosing the control command marked by a green check mark at each control state, a control equivalent and resilient supervisor $OS$ is extracted, which is illustrated in Fig. \ref{fig:OS}. 

\section{Conclusions}
\label{sec:conclusions}

In this work, we investigate the problem of obfuscating supervisors against covert actuator attackers. By constructing the behavior-preserving structure to exactly encode all the control equivalent supervisors, we propose a sound and complete algorithm to generate all the obfuscated supervisors and show the studied problem is decidable. In the future works, we shall continue investigating supervisor obfuscation against more powerful attacks, e.g., covert sensor-actuator attacks, in more challenging scenarios, e.g., networked systems. 
 
\begin{appendices}
\section{Proof of Lemma III.1} 
\label{appendix: Lemma III.1} 
(If) Firstly, it can be checked that $L(P_{\Sigma_{o}}(G||S)) \subseteq L(P_{\Sigma_{o}}(G)||P_{\Sigma_{o}}(S)) = L(P_{\Sigma_{o}}(G)||S)$. Next, we prove that $L(G||S) \subseteq L(G||S')$. Thus, we need to show that for any $t \in L(G||S)$, we have $t \in L(G||S')$. Since $t \in L(G||S)$, we have $t \in L(G)$ and $t \in L(S)$. Thus, to prove $t \in L(G||S') = L(G) \cap L(S')$, we only need to show $t \in L(S')$. Since $t \in L(G||S) \subseteq L(P_{\Sigma_{o}}(G||S))$, we have $t \in L(P_{\Sigma_{o}}(G||S)) = L(P_{\Sigma_{o}}(G||S')) \subseteq L(P_{\Sigma_{o}}(G)||S') = L(P_{\Sigma_{o}}(G)) \cap L(S')$, which implies that $t \in L(S')$. Thus, $L(G||S) \subseteq L(G||S')$. By the same way, we could prove that $L(G||S') \subseteq L(G||S)$. Hence, $L(G||S) = L(G||S')$.

(Only if) The necessity is straightforward. \hfill $\blacksquare$

\section{Proof of Proposition III.1} 
\label{appendix: Proposition III.1} 
Since $L(BT(S')) \subseteq L(CE)$ and $L(BPNS) = L(BPS) \cap L(CE)$, to prove $L(BT(S')) \subseteq L(BPNS)$, we only need to show that $L(BT(S')) \subseteq L(BPS)$. Thus, we need to prove for any $t \in L(BT(S'))$, we have $t \in L(BPS)$. We adopt the mathematical induction to prove this result. The base case is: $t = \varepsilon$. Clearly, $\varepsilon \in L(BT(S'))$ and $\varepsilon \in L(BPS)$. Thus, the base case holds. Next, the induction hypothesis is that: for any $t \in L(BT(S'))$, we have $t \in L(BPS)$, when $|t| = k$. Then we shall show that for any $t\sigma \in L(BT(S'))$, we have $t\sigma \in L(BPS)$. For convenience, we denote $BT(S') = (Q_{bs'}, \Sigma \cup \Gamma, \xi_{bs'}, q_{bs'}^{init})$. According to the construction procedure of $BT(S')$, we have $L(BT(S')) \subseteq \overline{(\Gamma\Sigma_{uo}^{*}\Sigma_{o})^{*}}$, and then the verification can be divided into the following two cases:

1. $\sigma \in \Gamma$. For convenience, we shall denote $\sigma = \gamma \in \Gamma$. Based on the structure of $BT(S')$, we have $t \in (\Gamma\Sigma_{uo}^{*}\Sigma_{o})^{*}$. 
Then we have the following two subcases:
\begin{enumerate}[1)]
\setlength{\itemsep}{3pt}
\setlength{\parsep}{0pt}
\setlength{\parskip}{0pt}
    \item $\xi_{bps}(q_{bps}^{init}, t) = q^{dump}$. Based on Step 3.f in the construction of $BPS$, we have $En_{BPS}(\xi_{bps}(q_{bps}^{init}, t)) =  \Gamma \cup \Sigma$. Thus, it holds that $t\gamma \in L(BPS)$.
    \item $\xi_{bps}(q_{bps}^{init}, t) \neq q^{dump}$. Since $t \in (\Gamma\Sigma_{uo}^{*}\Sigma_{o})^{*}$, i.e., $t$ is ended with an event in $\Sigma_{o}$, based on the construction procedure of $BPS$ from $B$, we know that $\xi_{bps}(q_{bps}^{init}, t) \in Q_{b}^{com} \cup \{q^{dump}\}$. In addition, since $\xi_{bps}(q_{bps}^{init}, t) \neq q^{dump}$, we have $\xi_{bps}(q_{bps}^{init}, t) \in Q_{b}^{com}$. By construction of $BPS$, we have $P(t) \in L(B)$ and $(\xi_{b}(q_{b}^{init}, P(t)))^{com} = \xi_{bps}(q_{bps}^{init}, t)$, where $P: (\Sigma \cup \Gamma)^{*} \rightarrow \Sigma_{o}^{*}$.
    Then we show that at the state $\xi_{bps}(q_{bps}^{init}, t)$, the event $\gamma$ satisfies the conditions $\mathcal{C}_{1}$ and $\mathcal{C}_{2}$ presented in Step 3.a. For $\mathcal{C}_{1}$, it requires that $En_{B}(\xi_{b}(q_{b}^{init}, P(t))) \subseteq \gamma$. Since $L(G||S) = L(G||S')$, we have $L(B) = L(P_{\Sigma_{o}}(G||S)) = L(P_{\Sigma_{o}}(G||S')) \subseteq L(S')$. Thus, we have $En_{B}(\xi_{b}(q_{b}^{init}, P(t))) \subseteq En_{S'}(\xi_{s'}(q_{s'}^{init}, P(t)))$. Since $t\gamma \in L(BT(S'))$, we have $En_{S'}(\xi_{s'}(q_{s'}^{init}, P(t))) \\= \gamma$. Thus, $En_{B}(\xi_{b}(q_{b}^{init}, P(t))) \subseteq \gamma$. For $\mathcal{C}_{2}$, it requires that $(\forall (q_{g},q_{s}) \in \xi_{b}(q_{b}^{init}, P(t)))En_{G}(q_{g}) \cap En_{S'}(\xi_{s'}(q_{s'}^{init}, P(t))) \subseteq En_{B}(\xi_{b}(q_{b}^{init}, P(t)))$, which clearly holds; otherwise, we know that there exists $(q_{g},q_{s}) \in \xi_{b}(q_{b}^{init}, P(t))$ such that $En_{G}(q_{g}) \cap En_{S'}(\xi_{s'}(q_{s'}^{init}, P(t))) \not\subseteq En_{B}(\xi_{b}(q_{b}^{init}, P(t)))$, and then we have $L(P_{\Sigma_o}(G||S)) \neq L(P_{\Sigma_o}(G||S'))$, implying that
    $L(G||S) \neq L(G||S')$ based on \textbf{Lemma III.1}, which causes the contradiction.
\end{enumerate}
2. $\sigma \in \Sigma$. Based on the structure of $BT(S')$, there exists $t_{1} \in (\Gamma\Sigma_{uo}^{*}\Sigma_{o})^{*}$, $\gamma \in \Gamma$ and $t_{2} \in (\gamma \cap \Sigma_{uo})^{*}$ such that $t = t_{1}\gamma t_{2} \in L(BPS)$ and $\sigma \in \gamma$. Then we have the  following two subcases:
\begin{enumerate}[1)]
\setlength{\itemsep}{3pt}
\setlength{\parsep}{0pt}
\setlength{\parskip}{0pt}
    \item $\xi_{bps}(q_{bps}^{init}, t) = q^{dump}$. Based on Step 3.f in the construction of $BPS$, we have $En_{BPS}(\xi_{bps}(q_{bps}^{init}, t)) =  \Gamma \cup \Sigma$. Thus, it holds that $t\sigma \in L(BPS)$.
    \item $\xi_{bps}(q_{bps}^{init}, t) \neq q^{dump}$. Since any event in $\Sigma \cup \Gamma$ labels a self-loop transition at the state $q^{dump}$, we know that $\xi_{bps}(q_{bps}^{init}, t_{1}) \neq q^{dump}$. Thus, $\xi_{bps}(q_{bps}^{init}, t_{1}\gamma)$ is a reaction state. According to Step 3.b and Step 3.d in the construction of $BPS$, we have $\xi_{bps}(q_{bps}^{init}, t_{1}\gamma) = \xi_{bps}(q_{bps}^{init}, t_{1}\gamma t_{2})$, which is still a reaction state. In addition, according to Step 3.b - Step 3.e in the construction of $BPS$, any event in $\Sigma$ is defined at any reaction state, we have $t\sigma = t_{1}\gamma t_{2} \sigma \in L(BPS)$.
\end{enumerate}
Based on the above analysis, in any case, we have $t\sigma \in L(BPS)$, which completes the proof. \hfill $\blacksquare$ 

\section{Proof of Proposition III.2} 
\label{appendix: Proposition III.2} 
Since $L(G||S) \neq L(G||S')$, based on \textbf{Lemma III.1}, we have $L(B) = L(P_{\Sigma_{o}}(G||S)) \neq L(P_{\Sigma_{o}}(G||S')) = L(B')$, where $B' = P_{\Sigma_{o}}(G||S') = (Q_{b'}, \Sigma, \xi_{b'}, q_{b'}^{init})$. Then we know that there exists $t \in \Sigma_{o}^{*} \cap L(B) \cap L(B')$ such that for any $i \in [0: |t|-1]$, the following conditions are satisfied:
\begin{enumerate}[1)]
\setlength{\itemsep}{3pt}
\setlength{\parsep}{0pt}
\setlength{\parskip}{0pt}
    \item $En_{B}(\xi_{b}(q_{b}^{init}, \mathcal{P}_{i}(t))) = En_{B'}(\xi_{b'}(q_{b'}^{init}, \mathcal{P}_{i}(t)))$
    \item $En_{B}(\xi_{b}(q_{b}^{init}, t)) \neq En_{B'}(\xi_{b'}(q_{b'}^{init}, t))$
\end{enumerate}
According to the way of constructing $B$ and $B'$, we have for any $i \in [0: |t|-1]$, the following conditions are satisfied:
\begin{enumerate}[C1)]
\setlength{\itemsep}{3pt}
\setlength{\parsep}{0pt}
\setlength{\parskip}{0pt}
    \item $En_{B}(\xi_{b}(q_{b}^{init}, \mathcal{P}_{i}(t))) \subseteq En_{S'}(\xi_{s'}(q_{s'}^{init}, \mathcal{P}_{i}(t)))$
    \item $(\forall (q_{g}, q_{s}) \in \xi_{b}(q_{b}^{init}, \mathcal{P}_{i}(t)))En_{G}(q_{g}) \cap En_{S'}(\xi_{s'}(q_{s'}^{init}, \\ \mathcal{P}_{i}(t))) \subseteq En_{B}(\xi_{b}(q_{b}^{init}, \mathcal{P}_{i}(t)))$
    \item $En_{B}(\xi_{b}(q_{b}^{init}, t)) \not\subseteq En_{S'}(\xi_{s'}(q_{s'}^{init}, t)) \vee \\ (\exists (q_{g}, q_{s}) \in \xi_{b}(q_{b}^{init}, t))En_{G}(q_{g}) \cap En_{S'}(\xi_{s'}(q_{s'}^{init}, t)) \\ \not\subseteq En_{B}(\xi_{b}(q_{b}^{init}, t))$
\end{enumerate}
Next, we consider two strings $u = \gamma_{0}t[1]\gamma_{1}\dots t[|t|-1]\gamma_{|t|-1}t[|t|]$ ($u = \varepsilon$ if $t = \varepsilon$) and $u\gamma_{|t|}$, where for any $i \in [0: |t|]$, we have $\gamma_{i} = En_{S'}(\xi_{s'}(q_{s'}^{init}, \mathcal{P}_{i}(t)))$. Since $t \in \Sigma_{o}^{*} \cap L(B) \cap L(B')$, we know that $t \in L(S')$. Thus, for any $j \in [1: |t|]$, it holds that $t[j] \in \gamma_{j-1}$. In addition, according to the construction procedure of $BT(S')$, we know that $u \in L(BT(S'))$. Next, we prove that $u \in L(BPS)$ by mathematical induction. For convenience, we denote $u = c_{1}\dots c_{|t|}$, where $c_{i} = \gamma_{i-1}t[i]$. The base case is to prove $c_{1} = \gamma_{0}t[1] \in L(BPS)$. If $t = \varepsilon$, then $u = \varepsilon$, which means that $c_{1} = \varepsilon \in L(BPS)$. Next, we only consider $t \neq \varepsilon$. Since $t \in L(P_{\Sigma_{o}}(G||S))$, we have $t[1] \in L(P_{\Sigma_{o}}(G||S))$. In addition, since the condition C1) and C2) hold, we know that for Step 3.a in the construction procedure of $BPS$, the condition $\mathcal{C}_{1}$ and $\mathcal{C}_{2}$ are satisfied for $\gamma_{0}$ at the state $(q_{b}^{init})^{com}$ in $BPS$. Thus, $\gamma_{0}t[1] \in L(BPS)$ and the base case holds. The induction hypothesis is $c_{1}\dots c_{k} = \gamma_{0}t[1]\gamma_{1}\dots \gamma_{k-1}t[k] \in L(BPS)$ and we need to prove $c_{1}\dots c_{k+1} = \gamma_{0}t[1]\gamma_{1}\dots \gamma_{k-1}t[k]\gamma_{k}t[k+1] \in L(BPS)$, where the hypothesis holds for $k \leq |t|-2$. It can be checked that $BPS$ would transit to the state $(\xi_{b}(q_{b}^{init}, t[1]\dots t[k]))^{com}$ via the string $c_{1}\dots c_{k}$. Thus, we need to check whether the condition $\mathcal{C}_{1}$ and $\mathcal{C}_{2}$ in Step 3.a of the construction procedure of $BPS$ are satisfied for $\gamma_{k}$ at the state $(\xi_{b}(q_{b}^{init}, t[1]\dots t[k]))^{com}$, that is, whether $En_{B}(\xi_{b}(q_{b}^{init}, t[1]\dots t[k])) \subseteq \gamma_{k}$ and $(\forall (q_{g},q_{s}) \in \xi_{b}(q_{b}^{init}, t[1]\dots t[k]))En_{G}(q_{g}) \cap \gamma_{k} \subseteq En_{B}(\xi_{b}(q_{b}^{init}, t[1]\dots t[k]))$ hold. Clearly, these two conditions hold as it is a special case of C1) and C2) when $i = k$. Thus, $u \in L(BPS)$. Since $BPNS = BPS||CE$ and $t[j] \in \gamma_{j-1}$ ($j \in [1: |t|]$), we know that $u \in L(BPNS)$.

Finally, we prove that $u\gamma_{|t|} \in L(BT(S'))$ and $u\gamma_{|t|} \notin L(BPS)$. According to the way of generating $BT(S')$, we have $u\gamma_{|t|} \in L(BT(S'))$ because $\gamma_{|t|} = En_{S'}(\xi_{s'}(q_{s'}^{init}, t))$. Since the condition C3) holds, we know that for $BPS$, at the state $q^{com} = (\xi_{b}(q_{b}^{init}, t))^{com} = \xi_{bps}(q_{bps}^{init}, u)$, it holds that either $En_{B}(q) \not\subseteq En_{S'}(\xi_{s'}(q_{s'}^{init}, t)) = \gamma_{|t|}$ or $(\exists (q_{g}, q_{s}) \in q)En_{G}(q_{g}) \cap En_{S'}(\xi_{s'}(q_{s'}^{init}, t)) = En_{G}(q_{g}) \cap \gamma_{|t|} \not\subseteq En_{B}(q)$, i.e., the conditions $\mathcal{C}_{1}$ and $\mathcal{C}_{2}$ in Step 3.a of the construction procedure of $BPS$ are not satisfied for the state $q^{com} = \xi_{bps}(q_{bps}^{init}, u)$, rendering that $u\gamma_{|t|} \notin L(BPS)$ and thus $u\gamma_{|t|} \notin L(BPNS)$, which completes the proof.  \hfill $\blacksquare$

\section{Proof of Theorem III.1} 
\label{appendix: Theorem III.1} 
Based on \textbf{Proposition III.1}, we have LHS $\subseteq$ RHS. Next, we prove RHS $\subseteq$ LHS. Thus, we need to show that for any $t \in L(BPNS)$, we have $t \in$ LHS. We adopt the contradiction and assume that $t \notin$ LHS. Since $t \in L(BPNS) = L(BPS) \cap L(CE)$, we have $t \in L(BPS)$ and $t \in L(CE)$. In addition, since $CE$ encodes all the bipartite supervisors and $t \notin$ LHS, we know that there exists a supervisor $\hat{S}$ such that $L(G||S) \neq L(G||\hat{S})$ and $t \in L(BT(\hat{S})) -$ LHS. Then, $t$ must contain some control command that would result in the violation of control equivalence. Without loss of generality, we know that there exists $u \leq t$ such that $u = \gamma_{0}t_{1}\gamma_{1}\dots t_{m}\gamma_{m}$, where $m \in \mathbb{N}$ ($u = \gamma_{0}$ when $m = 0$) and the following conditions are satisfied:
\begin{enumerate}[1.]
\setlength{\itemsep}{3pt}
\setlength{\parsep}{0pt}
\setlength{\parskip}{0pt}
    \item $(\forall i \in [1:m])t_{i} \in (\gamma_{i-1} \cap \Sigma_{uo})^{*}(\gamma_{i-1} \cap \Sigma_{o})$ for $m \geq 1$. For convenience, we denote $t^{obs} = t_{1}^{\downarrow}\dots t_{m}^{\downarrow}$ for $m \geq 1$, and $t^{obs} = \varepsilon$ for $m = 0$.
    \item $(\forall i \in [1:m])En_{B}(\xi_{b}(q_{b}^{init}, \mathcal{P}_{i-1}(t^{obs}))) \subseteq \gamma_{i-1}$ for $m \geq 1$.
    \item $(\forall i \in [1:m])(\forall (q_{g}, q_{s}) \in \xi_{b}(q_{b}^{init}, \mathcal{P}_{i-1}(t^{obs})))En_{G}(q_{g})\\ \cap \gamma_{i-1} \subseteq En_{B}(\xi_{b}(q_{b}^{init}, \mathcal{P}_{i-1}(t^{obs})))$ for $m \geq 1$.
    \item $En_{B}(\xi_{b}(q_{b}^{init}, t^{obs})) \not\subseteq \gamma_{m} \vee (\exists (q_{g}, q_{s}) \in \xi_{b}(q_{b}^{init}, t^{obs}))\\En_{G}(q_{g}) \cap \gamma_{m} \not\subseteq En_{B}(\xi_{b}(q_{b}^{init}, t^{obs}))$
\end{enumerate}
Based on the above item 4, we know that $\mathcal{C}_{1} \wedge \mathcal{C}_{2}$ is not satisfied for the control command $\gamma_{m}$ at the state $(\xi_{b}(q_{b}^{init}, t^{obs}))^{com}$ in Case 3.a of the construction of $BPS$. Thus, $u \notin L(BPS)$ and $t \notin L(BPS)$, which causes the contradiction. Hence, the assumption does not hold and $t \in$ LHS, which completes the proof. \hfill $\blacksquare$

\section{Proof of Proposition IV.1} 
\label{appendix: Proposition IV.1} 
We denote $BT(S') = (Q_{bs'}, \Sigma \cup \Gamma, \xi_{bs'}, q_{bs'}^{init})$ and $BT(S')^{A} = (Q_{bs'}^{a}, \Sigma \cup \Gamma, \xi_{bs'}^{a}, q_{bs'}^{a,init})$. To prove $L(BT(S')^{A}) \subseteq L(BPNS^{A})$, we only need to demonstrate that $BT(S')^{A}$ is simulated by $BPNS^{A}$. Let $R \subseteq Q_{bs'}^{a} \times Q_{bpns}^{a} = (Q_{bs'} \cup \{q^{detect}\}) \times (Q_{bpns} \cup \{q_{bpns}^{detect}\})$ be a relation defined such that
\begin{enumerate}[1.]
\setlength{\itemsep}{3pt}
\setlength{\parsep}{0pt}
\setlength{\parskip}{0pt}
    \item For any $q_{1} \in Q_{bs'} \subseteq Q_{bs'}^{a}$, any $q_{2} \in Q_{bpns} \subseteq Q_{bpns}^{a}$ and any $t \in L(BT(S')) \subseteq L(BPNS)$ such that $\xi_{bs'}(q_{bs'}^{init}, t) = q_{1}$ and $\xi_{bpns}(q_{bpns}^{init}, t) = q_{2}$, $(q_{1}, q_{2}) \in R$
    \item $(q^{detect}, q_{bpns}^{detect}) \in R$
\end{enumerate}
We observe that, by construction, $(q_{bs'}^{a,init}, q_{bpns}^{a,init}) \in R$. Next, without loss of generality, we consider two states $q_{1} \in Q_{bs'}$ and $q_{2} \in Q_{bpns}$ such that $(q_{1}, q_{2}) \in R$. According to the definition of $R$, we know that there exists $t \in L(BT(S')) \subseteq L(BPNS)$ such that $\xi_{bs'}(q_{bs'}^{init}, t) = q_{1}$ and $\xi_{bpns}(q_{bpns}^{init}, t) = q_{2}$. Since $L(BT(S')) \subseteq \overline{(\Gamma\Sigma_{uo}^{*}\Sigma_{o})^{*}}$, there are three cases:
\begin{enumerate}[1.]
\setlength{\itemsep}{3pt}
\setlength{\parsep}{0pt}
\setlength{\parskip}{0pt}
    \item $t \in (\Gamma\Sigma_{uo}^{*}\Sigma_{o})^{*}\Gamma$. We know that $q_{1}$ and $q_{2}$ are reaction states and only events in $\Sigma$ are defined at $q_{1}$ and $q_{2}$. Then, for any $\sigma \in \Sigma$ such that $\xi_{bs'}^{a}(q_{1}, \sigma) = \hat{q}_{1}$, we have the following analysis: Firstly, by construction, it can be checked that $En_{BT(S')}(q_{1}) = En_{BPNS}(q_{2})$. Secondly, as presented in Step 3.b-Step 3.c of the construction of $BT(S')^{A}$ and Step 3.b-Step 3.c of the construction of $BPNS^{A}$, we know that 1) the new transitions in $BT(S')^{A}$ ($BPNS^{A}$, respectively) would only be added at the reaction states of $BT(S')$ ($BPNS$, respectively), and 2) at any reaction state in $BT(S')$ ($BPNS$, respectively), we shall complete the undefined events in $(\Sigma_{c,a} \cap \Sigma_{uo}) \cup \Sigma_{o}$, where the unobservable events are self-loop transitions and the observable events lead to the state $q^{detect}$ ($q_{bpns}^{detect}$, respectively). Thus, $\xi_{bpns}^{a}(q_{2}, \sigma)!$ and we denote $\xi_{bpns}^{a}(q_{2}, \sigma) = \hat{q}_{2}$. If $\sigma \in En_{BT(S')}(q_{1})$, then we have $\xi_{bs'}(q_{bs'}^{init}, t\sigma) = \hat{q}_{1}$ and $\xi_{bpns}(q_{bpns}^{init}, t\sigma) = \hat{q}_{2}$, i.e., $(\hat{q}_{1}, \hat{q}_{2}) \in R$. If $\sigma \notin En_{BT(S')}(q_{1})$, then we have the following two subcases: 1) $\sigma \in \Sigma_{uo}$. Since unobservable events are self-loop transitions, we know that $\hat{q}_{1} = q_{1}$ and $\hat{q}_{2} = q_{2}$, i.e., $(\hat{q}_{1}, \hat{q}_{2}) \in R$. 2) $\sigma \in \Sigma_{o}$. Then we know that $\hat{q}_{1} = q^{detect}$ and $\hat{q}_{2} = q_{bpns}^{detect}$. In addition, since $(q^{detect}, q_{bpns}^{detect}) \in R$, we still have $(\hat{q}_{1}, \hat{q}_{2}) \in R$.
    \item $t \in (\Gamma\Sigma_{uo}^{*}\Sigma_{o})^{*}\Gamma\Sigma_{uo}^{*}$. Since unobservable events are self-loop transitions at any reaction state of $BT(S')^{A}$ and $BPNS^{A}$, this case can be reduced to Case 1.
    \item $t \in (\Gamma\Sigma_{uo}^{*}\Sigma_{o})^{*}$. We know that $q_{1}$ and $q_{2}$ are control states and only events in $\Gamma$ are defined at $q_{1}$ and $q_{2}$. Then, for any $\gamma \in \Gamma$ such that $\xi_{bs'}^{a}(q_{1}, \gamma) = \hat{q}_{1}$, firstly, we know that $t\gamma \in L(BT(S'))$. Based on \textbf{Proposition III.1}, we have $t\gamma \in L(BPNS)$, and thus, by construction, $t\gamma \in L(BPNS)^{A}$, i.e., $\xi_{bpns}^{a}(q_{2}, \sigma)!$ and we denote $\xi_{bpns}^{a}(q_{2}, \sigma) = \hat{q}_{2}$. Clearly, $(\hat{q}_{1}, \hat{q}_{2}) \in R$ as $\xi_{bs'}(q_{bs'}^{init}, t\gamma) = \hat{q}_{1}$ and $\xi_{bpns}(q_{bpns}^{init}, t\gamma) = \hat{q}_{2}$
\end{enumerate}
Thus, for any $\sigma \in \Sigma \cup \Gamma$ such that $\xi_{bs'}^{a}(q_{1}, \sigma) = \hat{q}_{1}$, we have $\xi_{bpns}^{a}(q_{2}, \sigma) = \hat{q}_{2}$ and $(\hat{q}_{1}, \hat{q}_{2}) \in R$, which completes the proof. \hfill $\blacksquare$

\section{Proof of Theorem IV.1} 
\label{appendix: Theorem IV.1} 
Based on \textbf{Proposition IV.1}, we have LHS $\subseteq$ RHS. Next, we prove RHS $\subseteq$ LHS, that is, for any $t \in$ RHS, we need to show $t \in$ LHS. Then there are two cases:
\begin{enumerate}[1.]
\setlength{\itemsep}{3pt}
\setlength{\parsep}{0pt}
\setlength{\parskip}{0pt}
    \item $t \in L(BPNS)$. Based on \textbf{Theorem III.1}, we have $t \in \bigcup\limits_{S' \in \mathscr{S}_{e}(S)}L(BT(S'))$. Since the contrscution of $BT(S')^{A}$ does not remove any transition originally defined in $BT(S)$, we have $t \in $ LHS.
    \item $t \notin L(BPNS)$ but $t \in L(BPNS^{A})$. Then we need to prove $t \in$ LHS, i.e., for any $n \in [0: |t|]$, we have $\mathcal{P}_{n}(t) \in$ LHS.
    We shall adopt the mathematical induction. For the base case, it clearly holds as $\mathcal{P}_{0}(t) = \varepsilon \in$ LHS. The induction hypothesis is $\mathcal{P}_{k}(t) \in$ LHS, where the hypothesis holds for $k \leq |t|-2$, and we need to prove $\mathcal{P}_{k+1}(t) := \mathcal{P}_{k}(t)\sigma \in$ LHS. Then there are two subcases:
    \begin{enumerate}[a.]
    \setlength{\itemsep}{3pt}
    \setlength{\parsep}{0pt}
    \setlength{\parskip}{0pt}
        \item $\mathcal{P}_{k}(t) = t_{1}\gamma t_{2}$, where $t_{1} \in (\Gamma\Sigma_{uo}^{*}\Sigma_{o})^{*}$, $\gamma \in \Gamma$, $t_{2} \in (\gamma \cap \Sigma_{uo})^{*}$. Since $\mathcal{P}_{k}(t) \in$ LHS, there exists a supervisor $S' \in \mathscr{S}_{e}(S)$ such that $\mathcal{P}_{k}(t) \in L(BT(S')^{A})$ and $L(BT(S')^{A})$ would reach a reaction state $q$ via the string $\mathcal{P}_{k}(t)$. 
        We denote $BT(S')^{A} = (Q_{bs'}^{a}, \Sigma \cup \Gamma, \xi_{bs'}^{a}, q_{bs'}^{a,init})$. Since the construction of $BT(S')^{A}$ from $BT(S)$ follows the same operation as that of the construction of $BPNS^{A}$ from $BPNS$, we have $En_{BPNS^{A}}(\xi_{bpns}^{a}(q_{bpns}^{a,init}, \mathcal{P}_{k}(t))) = En_{BT(S')^{A}}(\xi_{bs'}^{a}(q_{bs'}^{a,init}, \mathcal{P}_{k}(t)))$. Thus, $\mathcal{P}_{k+1}(t) = \mathcal{P}_{k}(t)\sigma \in L(BT(S')^{A}) \subseteq$ LHS.
        \item $\mathcal{P}_{k}(t) \in (\Gamma\Sigma_{uo}^{*}\Sigma_{o})^{*}$. It can be checked that $BPNS^{A}$ does not reach the state $q_{bpns}^{detect}$ via $\mathcal{P}_{k}(t)$; otherwise, if $BPNS^{A}$ reaches the state $q_{bpns}^{detect}$ via $\mathcal{P}_{k}(t)$, since there are no events in $\Sigma \cup \Gamma$ defined at the state $q_{bpns}^{detect}$, then the string is halted at $\mathcal{P}_{k}(t)$, which causes the contradiction with the fact that $\mathcal{P}_{k+1}(t)= \mathcal{P}_{k}(t)\sigma$. In addition, we know that $\sigma \in \Gamma$ because $L(BPNS^{A}) \subseteq \overline{(\Gamma\Sigma_{uo}^{*}\Sigma_{o})^{*}}$. Since the construction of $BPNS^{A}$ does not remove from $BPNS$ any transition that is labelled by an event in $\Gamma$ and $\mathcal{P}_{k}(t) \in$ LHS, based on \textbf{Theorem III.1}, there exists a supervisor $S' \in \mathscr{S}_{e}(S)$ such that $\mathcal{P}_{k+1}(t) = \mathcal{P}_{k}(t)\sigma \in L(BT(S')^{A}) \subseteq$ LHS.
    \end{enumerate}    
\end{enumerate}
Based on the above analysis, we have $t \in$ LHS, which completes the proof. \hfill $\blacksquare$ 

\section{Proof of Proposition IV.2} 
\label{appendix: Proposition IV.2} 
We adopt the contradiction and assume that $L(G||CE^{A}||BT(S')^{A}||\mathcal{A}) \not\subseteq L(G||CE^{A}||BPNS^{A}||\hat{\mathcal{A}})$.
Since $L(G||CE^{A}||BT(S')^{A}||\mathcal{A}) \subseteq \overline{(\Gamma\Sigma_{uo}^{*}\Sigma_{o})^{*}}$ and $L(G||CE^{A}||BPNS^{A}||\hat{\mathcal{A}}) \subseteq \overline{(\Gamma\Sigma_{uo}^{*}\Sigma_{o})^{*}}$, it must be the case that there exists $t \in \overline{(\Gamma\Sigma_{uo}^{*}\Sigma_{o})^{*}}$ such that $t \in L(G||CE^{A}||BT(S')^{A}||\mathcal{A}) = L(G||CE^{A}||BT(S')^{A}) \cap L(\mathcal{A})$ and $t \notin L(G||CE^{A}||BPNS^{A}||\hat{\mathcal{A}}) = L(G||CE^{A}||BPNS^{A}) \cap L(\hat{\mathcal{A}})$. Thus, $t \in L(G||CE^{A}||BT(S')^{A})$. In addition, based on \textbf{Proposition IV.1}, we have $L(BT(S')^{A}) \subseteq L(BPNS^{A})$. Since the alphabets of $BT(S')^{A}$ and $BPNS^{A}$ are the same, we have $L(G||CE^{A}||BT(S')^{A}) \subseteq L(G||CE^{A}||BPNS^{A})$. Thus, $t \in L(G||CE^{A}||BPNS^{A})$, based on which we have $t \notin L(\hat{\mathcal{A}})$. Since $\hat{\mathcal{A}}$ is synthesized by treating $\mathcal{P}$ as the plant and $\mathcal{P}_{r}$ as the requirement, we have the following two cases:
\begin{enumerate}[1.]
\setlength{\itemsep}{3pt}
\setlength{\parsep}{0pt}
\setlength{\parskip}{0pt}
    \item $t \notin L(\mathcal{P}_{r})$. Since $t \in L(\mathcal{P})$, according to the construction of $\mathcal{P}_{r}$, we know that via the string $t$, $G||CE^{A}||BT(S')^{A}||\mathcal{A}$ would reach the state $(q, q_{ce}^{a}, q_{bs'}^{a}, q_{a})$, where $q_{bs'}^{a} = q^{detect}$, i.e., $\mathcal{A}$ is not a covert actuator attacker against the supervisor $S'$, which causes the contradiction.
    \item $t \in L(\mathcal{P}_{r})$. According to Step 3 of \textbf{Procedure 1}, we know that $L(\hat{\mathcal{A}})$ is the supremal controllable and normal sublanguage of $L(G||CE^{A}||BPNS^{A})$ w.r.t. $L(\mathcal{P}_{r})$. In addition, based on the construction of $BPNS^{A}$ and $CE^{A}$, we know that any transition that would lead to a state in $Q_{bad}$ of $\mathcal{P}$ must be labelled by
    an event in $\Sigma_{c,a} \cap \Sigma_{o}$, which is controllable by the actuator attacker. Since $t \in L(\mathcal{P}_{r})$ and $t \notin L(\hat{\mathcal{A}})$, we know that there exists $\hat{t} = \hat{t}'\gamma\sigma_{1}\sigma_{2}\dots\sigma_{n}\sigma_{o} \in L(\mathcal{P})$ such that the following conditions are satisfied:
    \begin{enumerate}[1)]
    \setlength{\itemsep}{3pt}
    \setlength{\parsep}{0pt}
    \setlength{\parskip}{0pt}
        \item $(\exists u \leq t)P_{\Sigma_{o,a} \cup \Gamma}(\hat{t}) = P_{\Sigma_{o,a} \cup \Gamma}(u)$, where $P_{\Sigma_{o,a} \cup \Gamma}: (\Sigma \cup \Gamma)^{*} \rightarrow (\Sigma_{o,a} \cup \Gamma)^{*}$
        \item $\xi_{\mathcal{P}}(q_{\mathcal{P}}^{init}, \hat{t}) \in Q_{bad}$
        \item $\hat{t}^{\downarrow} = \sigma_{o} \in (\Sigma_{c,a} - \gamma) \cap \Sigma_{o}$
        \item $(\forall i \in [1:n]) \sigma_{i} \in (\gamma \cup \Sigma_{c,a}) \cap \Sigma_{uo}$
    \end{enumerate}
    Since $P_{\Sigma_{o,a} \cup \Gamma}(\hat{t}) = P_{\Sigma_{o,a} \cup \Gamma}(u)$, we know that $u = t'\gamma\sigma_{1}'\sigma_{2}'\dots\sigma_{m}'\sigma_{o}$ such that the following conditions are satisfied:
    \begin{enumerate}[1)]
    \setlength{\itemsep}{3pt}
    \setlength{\parsep}{0pt}
    \setlength{\parskip}{0pt}
        \item $P_{\Sigma_{o,a} \cup \Gamma}(t') = P_{\Sigma_{o,a} \cup \Gamma}(\hat{t}')$
        \item $(\forall i \in [1:m]) \sigma_{i}' \in (\gamma \cup \Sigma_{c,a}) \cap \Sigma_{uo}$
    \end{enumerate}    
    Since $\sigma_{o} \in (\Sigma_{c,a} - \gamma) \cap \Sigma_{o}$, we know that $BPNS^{A}$ would reach the state $q_{bpns}^{detect}$ via the string $u$, implying that $t = u$. Thus, $\xi_{\mathcal{P}}(q_{\mathcal{P}}^{init}, t) \in Q_{bad}$, which causes the contradiction with the fact that $t \in L(\mathcal{P}_{r})$.
\end{enumerate}
Based on the above analysis, the assumption $L(G||CE^{A}||\\BT(S')^{A}||\mathcal{A}) \not\subseteq L(G||CE^{A}||BPNS^{A}||\hat{\mathcal{A}})$ does not hold, which completes the proof. 
\hfill $\blacksquare$

\section{Proof of Theorem IV.2} 
\label{appendix: Theorem IV.2} 
Based on \textbf{Corollary IV.1}, we have LHS $\subseteq$ RHS. Then we only need to show RHS $\subseteq$ LHS. We shall adopt the contradiction and assume that RHS $\not\subseteq$ LHS. Then we know that there exists $\varepsilon \neq t \in \overline{(\Gamma\Sigma_{uo}^{*}\Sigma_{o})^{*}}$ such that $t \in$ RHS and $t \notin$ LHS. Hence, $t$ can be executed in $G$, $CE^{A}$, $BPNS^{A}$ and $\hat{\mathcal{A}}$, after we lift their alphabets to $\Sigma \cup \Gamma$, and $G$ would reach the state in $Q_{d}$ via $t$ after alphabet lift. Then, based on \textbf{Theorem IV.1}, we know that there exists $S' \in \mathscr{S}_{e}(S)$ such that $t \in L(BT(S')^{A})$. Thus, $t \in L_{m}(G||CE^{A}||BT(S')^{A}||\hat{\mathcal{A}})$, i.e., $\hat{\mathcal{A}}$ is damage-reachable against $S'$. In addition, it can be checked that $\hat{\mathcal{A}}$ is covert against $S'$; otherwise, there exists a string $t' \in L(G||CE^{A}||BT(S')^{A}||\hat{\mathcal{A}})$ such that $BT(S')^{A}$ reaches the state $q^{detect}$ via the string $t'$, which results in that $L(G||CE^{A}||BPNS^{A}||\hat{\mathcal{A}})$ reaches some state in $Q_{bad}$ via the string $t'$ and the contradiction is caused. Thus, $\hat{\mathcal{A}}$ is covert and damage-reachable against $S'$ and we have $\hat{\mathcal{A}} \in \mathscr{A}(S')$, which means that that $t \in$ LHS and this causes the contradiction. Hence, the assumption that RHS $\not\subseteq$ LHS does not hold, which completes the proof. \hfill $\blacksquare$

\section{Proof of Proposition IV.3} 
\label{appendix: Proposition IV.3} 
Firstly, we prove $L(S_{0}) \subseteq L(BPNS)$. We adopt the contradiction and assume that $L(S_{0}) \not\subseteq L(BPNS)$. Since $L(S_{0}) \subseteq \overline{(\Gamma\Sigma_{uo}^{*}\Sigma_{o})^{*}}$ and $L(BPNS) \subseteq \overline{(\Gamma\Sigma_{uo}^{*}\Sigma_{o})^{*}}$, we have the following two cases: 
\begin{enumerate}[1.]
\setlength{\itemsep}{3pt}
\setlength{\parsep}{0pt}
\setlength{\parskip}{0pt}
    \item There exists 
    $t \in L(S_{0}) \cap L(BPNS)$ and $\gamma \in \Gamma$ such that $t\gamma \in L(S_{0})$ and $t\gamma \notin L(BPNS)$. Thus, $t\gamma \in L(S_{0}) \subseteq L(S_{0}^{A}) \subseteq L(BPNS^{A})$. Since the construction of $BPNS^{A}$ from $BPNS$ does not add any transition labelled by a control command, we have $t\gamma \in L(BPNS)$, which causes the contradiction.
    \item There exists $t \in (\Sigma \cup \Gamma)^*$, $\gamma \in \Gamma$, $t' \in (\gamma \cap \Sigma_{uo})^{*}$ and $\sigma \in \Sigma$ such that $t\gamma t' \in L(S_{0}) \cap L(BPNS)$, $t\gamma t'\sigma \in L(S_{0})$ and $t\gamma t'\sigma \notin L(BPNS)$. Thus, $t\gamma t'\sigma \in L(S_{0}) \subseteq L(S_{0}^{A}) \subseteq L(BPNS^{A})$. Based on the construction of $BPNS^{A}$ from $BPNS$, we have $\sigma \in \Sigma - \gamma$. However, this would violate the structure of $S_{0}$, which causes the contradiction. 
\end{enumerate}
Thus, the assumption does not hold. It follows that $L(S_{0}) \subseteq L(BPNS)$. Hence, $L(ONS) \subseteq L(S_{0}) \subseteq L(BPNS)$. \hfill $\blacksquare$

\section{Proof of Theorem IV.3} 
\label{appendix: Theorem IV.3} 
Firstly, we prove LHS $\subseteq$ RHS. Thus, we shall show for any $S' \in \mathscr{S}_{e}^{r}(S)$, we have $L(BT(S')) \subseteq L(ONS)$. Based on \textbf{Proposition IV.1}, we have $L(BT(S')^{A}) \subseteq L(BPNS^{A})$. Next, we prove $L(BT(S')^{A}) \subseteq L(S_{0}^{A})$. We adopt the contradiction and assume that $L(BT(S')^{A}) \not\subseteq L(S_{0}^{A})$. Firstly, since $S'$ is a resilient supervisor, based on \textbf{Theorem IV.2}, we have $(\forall t \in L(BT(S')^{A}))t \notin L_{m}(\mathcal{P}) = L_{m}(G||CE^{A}||BPNS^{A}||\hat{\mathcal{A}})$; otherwise, there exists $t \in L(BT(S')^{A})$ such that $t \in L_{m}(\mathcal{P})$, i.e., there exists a covert and damage-reachable actuator attacker against $S'$, which is contradictory to the fact that $S'$ is resilient. Then, due to the assumption $L(BT(S')^{A}) \not\subseteq L(S_{0}^{A})$ and Step 3 of \textbf{Procedure 2} to synthesize $S_{0}^{A}$, we know that there exists $u \in L(BPNS^{A})$, $\gamma \in \Gamma$ and $v \in (\gamma \cup \Sigma_{c,a})^{*} - \{\varepsilon\}$ such that 
\[
\begin{aligned}
u\gamma v \in L_{m}(\mathcal{P}) \wedge [(&\exists t \in L(BT(S')^{A} ))t\gamma \in L(BT(S')^{A}) \wedge \\ &t\gamma \notin L(S_{0}^{A}) \wedge P_{\Sigma_{o} \cup \Gamma}(t) = P_{\Sigma_{o} \cup \Gamma}(u)]
\end{aligned}
\]
Since 1) $L(BPNS^{A}) \subseteq \overline{(\Gamma\Sigma_{uo}^{*}\Sigma_{o})^{*}}$ and $L(BT(S')^{A}) \subseteq \overline{(\Gamma\Sigma_{uo}^{*}\Sigma_{o})^{*}}$, where any event in $\Sigma_{uo}$, if defined, leads to a self-loop transition and any event in $\Sigma_{o} \cup \Gamma$, if defined, leads to an outgoing transition, and 2) $P_{\Sigma_{o} \cup \Gamma}(t) = P_{\Sigma_{o} \cup \Gamma}(u)$, we know that $u\gamma \in L(BT(S')^{A})$. Then, according to the construction procedure of $BT(S')^{A}$, we have $u\gamma v \in L(BT(S')^{A})$, which causes the contradiction because $u\gamma v \in L_{m}(\mathcal{P})$. Thus, the assumption does not hold and $L(BT(S')^{A}) \subseteq L(S_{0}^{A})$. Then, we have $L(BT(S')) \subseteq L(S_{0})$; otherwise, there exists a string $w \in L(BT(S')) \subseteq L(BT(S')^{A}) \subseteq L(S_{0}^{A})$ such that $w \notin L(S_{0})$, thus, based on the construction of $S_{0}^{A}$ from $S_{0}$, there exists $w', w'' \in \overline{\{w\}}$, $\gamma' \in \Gamma$ and $\sigma \in \Sigma - \gamma'$ such that $w' = w''\gamma'\sigma$, which violates the structure of $BT(S')$.  
Next, we prove $L(BT(S')) \subseteq L(ONS)$. We adopt the contradiction and assume that $L(BT(S')) \not\subseteq L(ONS)$. Then, according to \textbf{Procedure 3}, without loss of generality, we know that there exists $k \geq 0$ such that $L(BT(S')) \subseteq L(S_{k})$ and $L(BT(S')) \not\subseteq L(S_{k+1})$. Then we know that there exists $u' \in L(S_{k})$, $\gamma'' \in \Gamma$ and $\sigma' \in \gamma \cap \Sigma_{o}$ such that 
\[
\begin{aligned}
&En_{S_{k}}(\xi_{S_{k}}(q_{S_{k}}^{init}, u'\gamma'' \sigma')) = \varnothing \wedge
\\& [(\exists t' \in L(BT(S'))) t'\gamma'' \in L(BT(S')) \wedge t'\gamma'' \notin L(S_{k+1}) \wedge \\& P_{\Sigma_{o} \cup \Gamma}(u') = P_{\Sigma_{o} \cup \Gamma}(t')]
\end{aligned}
\]
Similarly, we have $u'\gamma'' \in L(BT(S'))$, and thus $u'\gamma'' \sigma' \in L(BT(S'))$. Since $L(BT(S')) \subseteq L(S_{k})$ and there is always a control command in $\Gamma$ defined at any control state of $BT(S')$, we have $En_{S_{k}}(\xi_{S_{k}}(q_{S_{k}}^{init}, u'\gamma'' \sigma')) \neq \varnothing$, which causes the contradiction. Thus, the assumption does not hold and $L(BT(S')) \subseteq L(ONS)$.

Secondly, we prove RHS $\subseteq$ LHS. Thus, we need to show for any $t \in$ RHS, we have $t \in$ LHS. Next, we shall construct a bipartite supervisor whose closed behavior contains the string $t$, and then we prove it is resilient and control equivalent to $BT(S)$. Firstly, we generate an automaton $T$ that recognizes $t$, i.e., $L_{m}(T) = t$. Then we compute its subset construction $P_{\Sigma_{o} \cup \Gamma}(T) = (Q_{t}, \Sigma \cup \Gamma, \xi_{t}, q_{t}^{init})$. By construction, we could denote $Q_{t} = Q_{t}^{rea} \dot{\cup } Q_{t}^{com}$, where $Q_{t}^{rea}$ is the set of reaction states and $Q_{t}^{com}$ is the set of control states. Then we shall complete some transitions in $P_{\Sigma_{o} \cup \Gamma}(T)$ and generate a new automaton $NC = (Q_{nc}, \Sigma \cup \Gamma, \xi_{nc}, q_{nc}^{init})$, which contains the necessary control command sequence encoded in $P_{\Sigma_{o} \cup \Gamma}(T)$. The construction procedure of $NC$ is given as follows.
\begin{enumerate}[1.]
\setlength{\itemsep}{3pt}
\setlength{\parsep}{0pt}
\setlength{\parskip}{0pt}
    \item $Q_{nc} = Q_{t} \cup \{q^{obs}\} \cup \{q^{\gamma}|\gamma \in \Gamma\}$
    \item \begin{enumerate}[a.]
        \setlength{\itemsep}{3pt}
        \setlength{\parsep}{0pt}
        \setlength{\parskip}{0pt}
        \item $(\forall q, q' \in Q_{t})(\forall \sigma \in \Sigma \cup \Gamma)\xi_{t}(q, \sigma) = q' \Rightarrow \xi_{nc}(q, \sigma) = q'$
        \item $(\forall q \in Q_{t}^{rea})(\forall \sigma \in \Sigma_{uo})\neg \xi_{t}(q, \sigma)! \Rightarrow \xi_{nc}(q, \sigma) = q$
        \item $(\forall q \in Q_{t}^{rea})(\forall \sigma \in \Sigma_{o})\neg \xi_{t}(q, \sigma)! \Rightarrow \xi_{nc}(q, \sigma) = q^{obs}$
        \item $(\forall q \in Q_{t}^{com})En_{P_{\Sigma_{o} \cup \Gamma}(T)}(q) = \varnothing \Rightarrow (\forall \gamma \in \Gamma)\xi_{nc}(q, \gamma) = q^{\gamma}$
        \item $(\forall \gamma \in \Gamma)\xi_{nc}(q^{obs}, \gamma) = q^{\gamma}$
        \item $(\forall \gamma \in \Gamma)(\forall \sigma \in \gamma \cap \Sigma_{o}) \xi_{ce}(q^{\gamma}, \sigma) = q^{obs}$.
        \item $(\forall \gamma \in \Gamma)(\forall \sigma \in \gamma \cap \Sigma_{uo}) \xi_{ce}(q^{\gamma}, \sigma) = q^{\gamma}$. 
    \end{enumerate}
    \item $q_{nc}^{init} = q_{t}^{init}$
\end{enumerate}
Briefly speaking, all the transitions in $P_{\Sigma_{o} \cup \Gamma}(T)$ are retained, denoted by Step 2.a. At any reaction state $q \in Q_{t}^{rea}$, all the undefined unobservable (observable, respectively) events in $P_{\Sigma_{o} \cup \Gamma}(T)$ are completed, which would lead to a self-loop transition (lead to the newly added state $q^{obs}$, respectively), denoted by Step 2.b (Step 2.c, respectively). At any reaction state $q \in Q_{t}^{com}$, if there are no control commands defined in $P_{\Sigma_{o} \cup \Gamma}(T)$, then we shall complete the transition labelled as any control command $\gamma$, which would lead to the state $q^{\gamma}$, denoted by Step 2.d. Finally, for the state $q^{obs}$ and $q^{\gamma}$ ($\gamma \in \Gamma$), we shall follow the same construction procedure as that of $CE$ to encode the command-event execution phase, denoted by Step 2.e - Step 2.g. Then we compute $NCS = NC||ONS = (Q_{ncs}, \Sigma \cup \Gamma, \xi_{ncs}, q_{ncs}^{init})$. By construction, we denote $Q_{ncs} = Q_{ncs}^{rea} \dot{\cup} Q_{ncs}^{com}$, where $Q_{ncs}^{rea}$ is the set of reaction states and $Q_{ncs}^{com}$ is the set of control states. Based on $NCS$, we shall generate a bipartite supervisor, denoted as $BT = (Q_{bt}, \Sigma \cup \Gamma, \xi_{bt}, q_{bt}^{init})$, where
\begin{enumerate}[1.]
\setlength{\itemsep}{3pt}
\setlength{\parsep}{0pt}
\setlength{\parskip}{0pt}
    \item $Q_{bt} = Q_{ncs} = Q_{ncs}^{rea} \dot{\cup} Q_{ncs}^{com}$
    \item \begin{enumerate}[a.]
        \setlength{\itemsep}{3pt}
        \setlength{\parsep}{0pt}
        \setlength{\parskip}{0pt}
        \item $(\forall q, q' \in Q_{bt})(\forall \sigma \in \Sigma)\xi_{ncs}(q, \sigma) = q' \Rightarrow \xi_{bt}(q, \sigma) = q'$
        \item For any control state $q \in Q_{ncs}^{com}$, we randomly pick a control command $\gamma \in En_{NCS}(q)$ and define that: for any reaction state $q' \in Q_{ncs}^{rea}$, if $\xi_{ncs}(q, \gamma) = q'$, then $\xi_{bt}(q, \gamma) = q'$ and for any control command $\gamma' \in En_{NCS}(q) - \{\gamma\}$, we have $\neg\xi_{bt}(q, \gamma')!$.
    \end{enumerate}
    \item $q_{bt}^{init} = q_{ncs}^{init}$
\end{enumerate}
Finally, we generate the automaton $Ac(BT)$. For convenience, we shall still denote $Ac(BT)$ as $BT$. Next, we prove that $BT$ is a control equivalent and resilient bipartite supervisor, and $t \in L(BT)$. We have the following two facts for $ONS$: 1) at any control state, there is at least one control command defined, which would lead to a reaction state, and 2) at any reaction state which is reached from a control state by a transition labelled as a control command $\gamma$, all the events in $\gamma$ are defined and any unobservable event would lead to a self-loop and any observable event would lead to a control state. Based on the construction of $NC$, it can be checked that the facts 1) and 2) also hold for $NC$. Since $NCS = NC||ONS$, the facts 1) and 2) also hold for $NCS$. According to the construction of $BT$, where we only define one control command at each control state, we know that the fact 2) holds for $BT$, and there is only one control command defined at any control state of $BT$. Thus, $BT$ is consistent with a bipartite supervisor structure. 

Next, firstly, we prove $BT$ is control equivalent to $S$. We adopt the contradiction and assume that $BT$ is not control equivalent to $S$. Based on \textbf{Proposition III.2}, we have $L(BT) \not\subseteq L(BPNS)$. Since $BT = NC||ONS$, we have $L(BT) \subseteq L(ONS)$. Based on \textbf{Proposition IV.3}, we have $L(BT) \subseteq L(ONS) \subseteq L(BPNS)$, which causes the contradiction. Hence, the assumption does not hold and $BT$ is control equivalent to $S$.

Secondly, we prove $BT$ is resilient. We adopt the contradiction and assume $BT$ is not resilient. We denote the version of $BT$ under attack as $BT^{A}$, whose construction procedure is given in Section \ref{subsubsec:Supervisor}. Clearly, we have $L(BT^{A}) \subseteq L(S_{0}^{A})$ as $L(BT) \subseteq L(ONS) \subseteq L(S_{0})$. Since $BT$ is not resilient, we know that there exists an attacker $\mathcal{A} \in \mathscr{A}(BT)$ and a string $t \in L(BT^{A}) \subseteq L(S_{0}^{A})$ such that $t \in L_{m}(G||CE^{A}||BT^{A}||\mathcal{A})$. Based on \textbf{Proposition IV.2}, we have $t \in L_{m}(\mathcal{P}) = L_{m}(G||CE^{A}||BPNS^{A}||\hat{\mathcal{A}})$. Due to the synthesis procedure in Step 3 of \textbf{Procedure 2}, we know that $t \notin L(S_{0}^{A})$, which causes the contradiction. Thus, the assumption does not hold and $BT$ is resilient.

Finally, we show that $t \in L(BT)$. By construction, we know that $t \in L(NC)$. Since $t \in$ RHS = $L(ONS)$ and $NCS = NC||ONS$, we have $t \in L(NCS)$. We adopt the contradiction and assume that $t \notin L(BT)$. In the construction of $BT$ from $NCS$, where we only remove control commands for those control states in $NCS$ where more than one control command is defined. Thus, we know that there exists $t' \leq t$ and $\gamma \in \Gamma$ such that 1) $t'\gamma \leq t$, 2) $|En_{NCS}(\xi_{ncs}(q_{ncs}^{init}, t'))| \geq 2$, and 3) we do not pick the control command $\gamma$ at the control state $\xi_{ncs}(q_{ncs}^{init}, t')$ when we construct $BT$. However, due to the construction of $NC$ and $NCS$, there is only one control command defined at the state $\xi_{ncs}(q_{ncs}^{init}, t')$, which means that we are supposed to retain the control command $\gamma$ at the state $\xi_{ncs}(q_{ncs}^{init}, t')$, and this cause the contradiction. Thus, the assumption does not hold and $t \in L(BT)$.

Based on the above analysis, $BT \in \mathscr{S}_{e}^{r}(S)$, and $t \in L(BT) \subseteq$ LHS. Thus, RHS $\subseteq$ LHS, which completes the proof. \hfill $\blacksquare$

\section{Proof of Proposition IV.4} 
\label{appendix: Proposition IV.4} 
Firstly, we have the following two facts: 1) at any reachable control state, only one control command in $\Gamma$ is defined, and such a transition would lead to a reaction state, and 2) at any reachable reaction state, which is reached from a control state via a transition labelled by $\gamma \in \Gamma$, all the events in $\gamma$ are defined, and any event in $\gamma \cap \Sigma_{uo}$ is a self-loop transition and any event in $\gamma \cap \Sigma_{o}$ would lead to a  control state. Thus, $OS$ is consistent with a bipartite supervisor structure. Secondly, we prove it is control equivalent to $S$. We adopt the contradiction and assume that $OS$ is not control equivalent to $S$. Based on \textbf{Proposition IV.3}, we have $L(OS) \subseteq L(ONS) \subseteq L(BPNS)$. Based on \textbf{Proposition III.2}, we have $L(OS) \not\subseteq L(BPNS)$, which causes the contradiction. Thus, the assumption does not hold and $OS$ is control equivalent to $S$. Thirdly, we prove that $OS$ is resilient. We adopt the contradiction and assume that $OS$ is not resilient. Then we know that there exists a covert and damage-reachable actuator attacker $\mathcal{A}$ such that $L_{m}(G||CE^{A}||OS^{A}||\mathcal{A}) \neq \varnothing$, where $OS^{A}$ is the attacked version of $OS$, whose construction procedure is presented in Section \ref{subsubsec:Supervisor}. Without loss of generality, we assume that $t \in L_{m}(G||CE^{A}||OS^{A}||\mathcal{A})$, which implies that $t \in L(OS^{A})$. Based on \textbf{Theorem IV.2}, we have $t \in L_{m}(\mathcal{P}) = L_{m}(G||CE^{A}||BPNS^{A}||\hat{\mathcal{A}})$. In addition, we have $L(OS^{A}) \subseteq L(S_{0}^{A})$ as $L(OS) \subseteq L(S_{0})$. Thus, $t \in L(OS^{A}) \subseteq L(S_{0}^{A})$. However, due to Step 3 of \textbf{Procedure 2}, we have $L_{m}(\mathcal{P}) \cap L(S_{0}^{A})  = \varnothing$, implying that $t \notin L(S_{0}^{A})$,   which causes the contradiction. Hence, the assumption does not hold and $OS$ is resilient, which completes the proof. \hfill $\blacksquare$

\end{appendices}

\end{document}